\documentclass[conference]{IEEEtran}
\usepackage{microtype}
\usepackage{soul}
\usepackage{subfigure}
\usepackage[dvipsnames]{xcolor}
\usepackage{lscape}
\usepackage{rotating,booktabs,makecell}
\usepackage{times}
\usepackage[small,bf]{caption}
\usepackage{paralist}
\usepackage{slashbox}

\usepackage{xspace}
\usepackage{mathtools}%
\usepackage{url}
\usepackage{graphicx}
\usepackage{latexsym}
\usepackage{amssymb}
\usepackage{amsfonts}
\usepackage{psfrag}
\usepackage{multirow}
\usepackage{wrapfig}
\usepackage{comment}
\usepackage[T1]{fontenc}
\usepackage{listings}
\usepackage{paralist}
\usepackage{epstopdf}
\usepackage{tablefootnote,footnote}
\usepackage{todonotes}
\usepackage{indentfirst}
\usepackage{verbatim}
\usepackage{tikz}
\usepackage{booktabs}
\usepackage[keeplastbox]{flushend}
\usetikzlibrary{shapes,snakes}
\usepackage{cite}
\usepackage{color, colortbl}
\definecolor{darkgreen}{RGB}{0,50,25}
\definecolor{vlightgray}{RGB}{240,240,240}

\usepackage[bookmarks=true,%
            bookmarksnumbered=true,%
            colorlinks=true,%
            linkcolor=linkcol,%
            citecolor=citecol,%
            urlcolor=urlcol,%
            hypertexnames=true,%
            pdfpagelabels]%
      {hyperref}
   \hypersetup{ pdfauthor = {Enrico Mariconti},
                pdftitle = {},
                pdfkeywords = {},
                pdfcreator = {},
                pdfproducer = {}
              }
   \definecolor{linkcol}{rgb}{0,0,0.5}
   \definecolor{citecol}{rgb}{0,0.5,0.3}
   \definecolor{urlcol}{rgb}{0.3,0,0}
\usepackage{url}
   \def\UrlFont{\sffamily\small}

\newif\ifcomment
\commenttrue 
\newif\ifwatermark
\watermarkfalse

\ifwatermark
      \usepackage{draftwatermark}
      \SetWatermarkScale{.7}
      \SetWatermarkText{\shortstack{In Submission:\\Do Not Distribute}}
      \SetWatermarkColor[rgb]{1,0.88,0.88}
\fi

\ifcomment
	\newcommand{\edc}[1]{\textbf{\em\color{red}#1}}
	\newcommand{\lucky}[1]{\textbf{\em\color{blue}#1}}
	\newcommand{\gs}[1]{\textbf{\em\color{purple}#1}}
	\newcommand{\enm}[1]{\textbf{\em\color{brown}#1}}
	\newcommand{\panos}[1]{\textbf{\em\color{orange}#1}}
\else  
    \newcommand\edc[1]{}
    \newcommand\lucky[1]{}
    \newcommand\gs[1]{}
    \newcommand\enm[1]{}
    \newcommand\panos[1]{}
\fi

\tikzstyle{highlighter} = [
  yellow,
  line width = \baselineskip,
]
\newcounter{highlight}[page]

\AtBeginShipout{\AtBeginShipoutUpperLeft{\ifthenelse{\value{highlight} > 0}{\tikz[remember picture, overlay]{\foreach \stroke in {1,...,\arabic{highlight}} \draw[highlighter] (begin highlight \stroke) -- (end highlight \stroke);}}{}}}

\DeclareRobustCommand{\ttfamily}{\fontfamily{lmtt}\selectfont}
\makeatletter
\input{t1lmtt.fd}
\@namedef{T1+lmtt}{}
\makeatother
\makeatletter
\newcommand\smallscriptsize{\@setfontsize\scriptsize{5.75}{6.75}}
\makeatother

\usepackage[marginal,hang]{footmisc}
\renewcommand{\footnoterule}{%
  \kern -3pt
  \hrule width 1in 
  \kern 2pt
}

\newcommand{\descr}[1]{\smallskip\noindent\textbf{#1}}

\makeatletter
\def\url@leostyle{%
  \@ifundefined{selectfont}{\def\UrlFont{}}%
  {\def\UrlFont{}}%
}
\makeatother	
\usepackage[hyphenbreaks]{breakurl}

\def\approach{\textsc{MaMaDroid}\xspace}
\def\droid{\textsc{DroidAPIMiner}\xspace}

\begin{document}
%


\IEEEoverridecommandlockouts
\title{\approach: Detecting Android Malware by Building Markov Chains of Behavioral Models$^*$\thanks{$^*$This paper appears in the Proceedings of 24th Network and Distributed System Security Symposium (NDSS 2017). Some experiments have been slightly updated in this version.}}
\sloppy
\author{Enrico Mariconti$^\dag$, Lucky Onwuzurike$^\dag$, Panagiotis Andriotis$^\ddag$,\\ Emiliano De Cristofaro$^\dag$, Gordon Ross$^\dag$, and Gianluca Stringhini$^\dag$ \\[0.5ex] 
$^\dag$University College London~~~$^\ddag$University of the West of England \vspace{1cm}}

\thispagestyle{empty}
\maketitle

\begin{abstract}
The rise in popularity of the Android platform has resulted in an explosion of
malware threats targeting it.
As both Android malware and the operating system itself constantly evolve, it is
very challenging to design robust malware mitigation techniques that can operate
for long periods of time without the need for modifications or costly
re-training. In this paper, we present \approach, an Android malware detection
system that relies on app behavior. \approach builds a behavioral model, in the
form of a Markov chain, from the sequence of abstracted API calls performed by an app,
and uses it to extract features and perform classification. By abstracting calls to their packages or families, \approach maintains resilience to API changes and keeps the feature set size manageable. 
We evaluate its accuracy on a dataset of 8.5K benign and 35.5K malicious apps
collected over a period of six years, showing that it not only effectively
detects malware (with up to 99\% F-measure), but also that the model built by
the system keeps its detection capabilities for long periods of time (on
average, 87\% and 73\% F-measure, respectively, one and two years after
training). Finally, we compare against \droid, a state-of-the-art system that
relies on the frequency of API calls performed by apps, showing that \approach significantly outperforms it.
\end{abstract}

\section{Introduction}
\label{sec:introduction}

In the first quarter of 2016, 85\% of smartphone sales were devices running Android~\cite{statistics}.
Due to its popularity, cybercriminals have increasingly targeted this ecosystem~\cite{androidtrend}, as malware running on mobile devices can be particularly lucrative -- e.g., allowing attackers to defeat two factor authentication~\cite{android2fa,google} or trigger leakage of sensitive information~\cite{gordon2015information}.
Detecting malware on mobile devices presents additional challenges compared to desktop/laptop computers: smartphones have limited battery life, making it infeasible to use traditional approaches
requiring constant scanning and complex computation~\cite{polakis2015powerslave}.
Therefore, Android malware detection is typically performed by Google in a centralized fashion, 
i.e., by analyzing apps submitted to the Play Store using a tool called
Bouncer~\cite{oberheide2012dissecting}. However, many malicious apps manage to avoid detection~\cite{trusted},
and anyway Android's openness enables manufacturers and users to install apps that
come from third-party market places, which might not perform any malware checks at all, or anyway not as accurately~\cite{Zhou2012hey}. 

As a result, the research community has devoted significant attention to malware detection on Android. %
Previous work has often relied on the permissions requested by apps~\cite{Enck2009, Sarma2012},
using models built from malware samples. This strategy, however, is prone to false positives, since there are often legitimate reasons for 
benign apps to request permissions classified as dangerous~\cite{Enck2009}.
Another approach, used by \droid~\cite{Aafer2013DroidAPIMiner}, is to perform  classification based on API calls frequently used by malware. However, relying on the most common calls observed during training prompts the need for constant retraining, due to the evolution of malware and the Android API alike. For instance, ``old'' calls are often deprecated with new API releases, so malware developers may switch to different calls to perform similar actions, which affects \droid's effectiveness due to its use of specific  calls. 

In this paper, we present a novel malware detection system for Android
that instead relies on the {\em sequence} of {\em abstracted} API calls performed by an app 
rather than their use or frequency, aiming to capture the behavioral model of the app.
Our system, which we call  \approach,  abstracts API calls to either the {\em package} name of the call (e.g., {\tt java.lang}) or its source (e.g.,
{\tt java}, {\tt android}, {\tt google}),  which we refer to as {\em family}.
Abstraction provides resilience to API changes in the Android framework as families and packages are added and removed 
less frequently than single API calls.
At the same time, this does not abstract away the behavior of an app: %
for instance, packages include classes and interfaces used to perform similar operations
on similar objects, so we can model the types of operations from the package name, independently
of the underlying classes and interfaces.
For example, we know that the {\tt java.io} package is used for system I/O and access to the file system, even though there are different classes and interfaces provided by the package for such operations.

After abstracting the calls, \approach analyzes the {\em sequence} of API calls performed by an app,
aiming to model the app's behavior. Our intuition is that malware may use calls for different operations, and in a different order,
than benign apps. For example, android.media.MediaRecorder can be used by any app that has permission to record audio, 
but the call sequence may reveal that malware only uses calls from this class {\em after} calls to getRunningTasks(), 
which allows recording conversations~\cite{zhang2015leave}, as opposed to benign apps where calls from the class may appear in {\em any} order. 
Relying on the sequence of abstracted calls allows us to model behavior in a more complex way
than previous work, which only looked at the presence or absence of certain API calls
or permissions~\cite{Aafer2013DroidAPIMiner,arp2014drebin}, while still keeping the problem
tractable~\cite{ComplMacLear}.
\approach builds a statistical model to represent the transitions between the API calls performed by an app,
specifically, we model these transitions as Markov chains, and use them to extract features 
and perform classification (i.e., labeling apps as benign
or malicious). Calls are abstracted to either their package or their family, i.e., \approach operates in
one of two modes, depending on the abstraction granularity.

We present a detailed evaluation of both classification accuracy (using F-measure, precision, and recall) and runtime performance of \approach, using a
dataset of almost 44K apps (8.5K benign and 35.5K malware samples). 
We include a mix of older and newer apps, from October 2010 to May 2016, 
verifying that our model is robust to changes in Android malware samples and APIs.
To the best of our knowledge,
this is the largest malware dataset used to evaluate an Android malware detection system in a
research paper. %
Our experimental analysis shows that \approach can effectively model both benign and malicious Android apps, and perform an
efficient classification on them. Compared to other systems such as \droid~\cite{Aafer2013DroidAPIMiner}, 
our approach allows us to account for changes in the Android
API, without the need to frequently retrain the classifier. 

We show that \approach is able to effectively detect unknown malware samples not only in the ``present,'' (with F-measure up to 99\%) but also consistently over the years
(i.e., when the system is trained on older samples and classification performed over newer ones), 
as it keeps an average detection accuracy, evaluated in terms of F-measure, of 87\% after one year and 73\% after two years (as opposed to
46\% and 42\% achieved by \droid~\cite{Aafer2013DroidAPIMiner}). We also highlight that when the system is not efficient anymore (when the test set is newer than the training set by more than two years), it is as a result of \approach having low recall, but maintaining high precision.
We also do the opposite, i.e., training on newer samples and verifying that the system can still detect old malware.
This is particularly important as it shows that \approach can detect newer threats, while still identifying malware samples that have been in the wild for some time.

\descr{Summary of Contributions.} First, we introduce a novel approach, implemented in a tool called \approach, to detect Android malware by abstracting API calls to their package and family, and using Markov chains to model the behavior of the apps through the sequences of API calls. Second, we can detect unknown samples on the same year of training with an F-measure of 99\%, but also years after training the system, meaning that \approach does not need continuous re-training. Our system is scalable as we model every single app independently from the others and can easily append app features in a new training set. Finally, compared to previous work~\cite{Aafer2013DroidAPIMiner}, \approach achieves significantly higher accuracy with reasonably fast running times, while also being more robust to evolution in malware development and changes in the Android API.

\noindent{\bf Paper Organization.} The rest of the paper is organized as follows. The next section presents the \approach system,
then, Section~\ref{sec:data} introduces the datasets used in our evaluation (Section~\ref{sec:evaluation}), while Section~\ref{sec:discussion} 
further discusses our results as well as its limitations. After reviewing related work in Section~\ref{sec:related}, the paper concludes in Section~\ref{sec:conclusion}.

\section{The MaMaDroid System}
\label{sec:method}

\subsection{Overview}
We now introduce \approach, a novel system for Android malware detection.
\approach characterizes the transitions between
different API calls performed by Android apps -- i.e., the sequence of API calls. 
It then models these transitions as Markov chains, which are in turn used to extract
features for machine learning algorithms to classify apps as benign or malicious. %
\approach does not actually use the sequence of {\em raw}
API calls, but abstracts each call to either its package or its family.
For instance, the API call getMessage() is parsed as:\vspace{-0.1cm}
$$ \small \underbrace{\overbrace{\underbrace{\mbox{java}}_\text{\bf family}\mbox{\hspace{-0.1cm}.lang}}^\text{\bf package}\mbox{\hspace{-0.05cm}.Throwable: String getMessage()}}_\text{\bf API call}\vspace{-0.15cm}$$

Given these two different types of abstractions, we have
two modes of operation for \approach, each using one of the types of abstraction. We test both, highlighting their advantages and 
disadvantages --- in a nutshell, the abstraction to family is more lightweight, while that to package is more fine-grained. %

\begin{figure}[t]
 \center
 \includegraphics[width=0.495\textwidth]{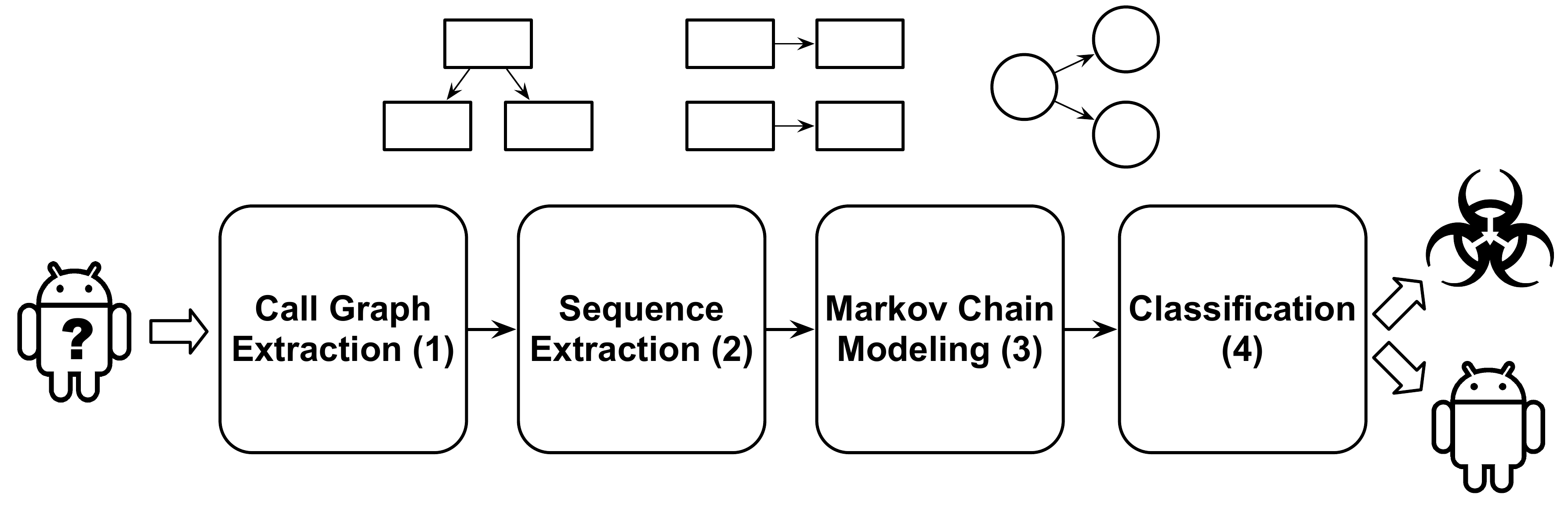}
 \caption{Overview of \approach operation. In (1), it extracts the call graph from an Android app, next, it builds the sequences of (abstracted) API calls from the call graph (2). In (3), the sequences of calls are used to build a Markov chain and a feature vector for that app. Finally, classification is performed in (4), labeling the app as benign or malicious.}
 \label{fig:diagram}
 \vspace{-0.15cm}
\end{figure}

\approach's operation goes through four phases, as depicted in
\figurename~\ref{fig:diagram}. First, we extract the call graph from each app by using static analysis (1), next
we obtain the sequences of API calls for the app using all unique nodes in the call graph and associating, to each node, all its
child nodes (2). As mentioned, we abstract a call to either its package
or family. Finally, by building on the sequences, \approach constructs a Markov
chain model (3), with the transition probabilities used as the feature vector to classify the app as either benign or malware using a machine learning classifier (4). In the rest of this section,
we discuss each of these steps in detail.

\subsection{Call Graph Extraction}\label{sec:extraction}
The first step in \approach is to extract the app's call graph.
We do so by performing static analysis on the app's apk.\footnote{The standard Android archive file format  
containing all files, including the Java bytecode, making up the app.}
Specifically, we use a Java optimization and analysis framework, 
Soot~\cite{ValleeRai1999Soot}, to extract call graphs %
and FlowDroid~\cite{Arzt2014flowdroid} to ensure contexts and flows are preserved. %

To better clarify the different steps involved in our system, we employ a ``running example,'' using a real-world malware sample.
Specifically, \figurename~\ref{lst:java} lists a class extracted from the decompiled apk of malware disguised as a memory booster app (with package name com.g.o.speed.memboost), which executes commands (rm, chmod, etc.) as root.\footnote{\url{https://www.hackread.com/ghost-push-android-malware/}} 
To ease presentation, we  focus on the portion of the code executed in the try/catch block.
The resulting call graph of the try/catch block is shown in \figurename~\ref{fig:calls}. Note that, for simplicity, 
we omit calls for object initialization, return types and parameters, as well as
implicit calls in a method. Additional calls that are invoked when getShell(true) is called are not
shown, except for the add() method that is directly called by the program code, as shown in \figurename~\ref{lst:java}. %

\begin{figure}[t]
\lstset{ %
breaklines=true,  
showspaces=false,
showstringspaces=false,
columns=flexible,
escapechar=@
}
\begin{lstlisting}[basicstyle=\ttfamily\smallscriptsize,language=Java]
package com.fa.c;

import android.content.Context;
import android.os.Environment;
import android.util.Log;
import com.stericson.RootShell.execution.Command;
import com.stericson.RootShell.execution.Shell;
import com.stericson.RootTools.RootTools;
import java.io.File;

public class RootCommandExecutor {
  public static boolean Execute(Context paramContext) {
    paramContext = new Command(0, new String[] { "cat " + Environment.getExternalStorageDirectory().getAbsolutePath() + File.separator + Utilities.GetWatchDogName(paramContext) + " > /data/" + Utilities.GetWatchDogName(paramContext), "cat " + Environment.getExternalStorageDirectory().getAbsolutePath() + File.separator + Utilities.GetExecName(paramContext) + " > /data/" + Utilities.GetExecName(paramContext), "rm " + Environment.getExternalStorageDirectory().getAbsolutePath() + File.separator + Utilities.GetWatchDogName(paramContext), "rm " + Environment.getExternalStorageDirectory().getAbsolutePath() + File.separator + Utilities.GetExecName(paramContext), "chmod 777 /data/" + Utilities.GetWatchDogName(paramContext), "chmod 777 /data/" + Utilities.GetExecName(paramContext), "/data/" + Utilities.GetWatchDogName(paramContext) + " " + Utilities.GetDeviceInfoCommandLineArgs(paramContext) + " /data/" + Utilities.GetExecName(paramContext) + " " + Environment.getExternalStorageDirectory().getAbsolutePath() + File.separator + Utilities.GetExchangeFileName(paramContext) + " " + Environment.getExternalStorageDirectory().getAbsolutePath() + File.separator + " " + Utilities.GetPhoneNumber(paramContext) });
    try {
      RootTools.getShell(true).add(paramContext);
      return true;
    }
    catch (Exception paramContext) {
     Log.d("CPS", paramContext.getMessage());
    }
    return false;
  }
}
\end{lstlisting}
\vspace{-0.25cm}
\caption{Code snippet from a malicious app (com.g.o.speed.memboost) executing commands as root.}
\vspace{-0.2cm}
\label{lst:java}
\end{figure}

\subsection{Sequence Extraction} \label{sec:sequence}
Next, we extract the sequences of API calls from the call graph. %
Since \approach uses static analysis, the graph obtained from Soot
represents the sequence of functions that are potentially called by the program.
However, each execution of the app could take a specific {\em branch} of the graph and only execute a
subset of the calls.
For instance, when running the code in \figurename~\ref{lst:java} multiple times,
the Execute method could be followed by different calls, e.g., getShell() in the try block only
or getShell() and then getMessage() in the catch block. %

In this phase, \approach operates as follows. First, it identifies a set of
entry nodes in the call graph, i.e., nodes with no incoming
edges (for example, the Execute method in the snippet from Fig.~\ref{lst:java} is the entry node if there is no incoming edge from any other call in the app). %
Then, it enumerates the paths reachable from each entry node. The
sets of all paths identified during this phase constitutes the sequences of API calls
which will be used to build a Markov chain behavioral model and
to extract features (see~Section~\ref{sec:MaCha}). %

\descr{Abstracting Calls to Families/Packages.} Rather than analyzing
raw API calls, we build \approach to work at a higher level, and operate in one of two modes by abstracting each call to either its package or family. 
This allows the system to be resilient to API changes and achieve scalability.
In fact, our experiments, presented in Section~\ref{sec:data}, show that, from a
dataset of 44K apps, we extract more than 10 million unique API calls, which
would result in a very large number of nodes, with the corresponding graphs (and feature vectors) being
quite sparse. Since as we will see the number of features used by \approach is the square of the number of nodes, having more than 10 million nodes would result in an impractical computational cost.

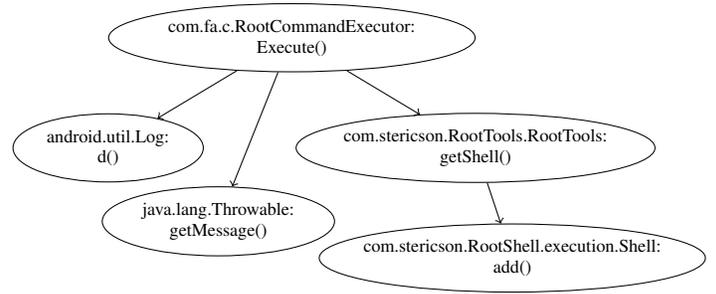
\begin{figure}[t!]
\begin{center}
\resizebox{.505\textwidth}{!}{\scriptsize
\begin{tikzpicture}[every node/.style={draw,ellipse},align=center]
\draw(0,0) node (A0) {com.fa.c.RootCommandExecutor:\\Execute()};
\draw(-2.5,-1.5) node (A1) {android.util.Log:\\ d()};
\draw(2.5,-1.5) node (A2) {\hspace*{-0.15cm}com.stericson.RootTools.RootTools:\hspace*{-0.15cm}\\ getShell()};
\draw(-1,-2.5) node (A3) {java.lang.Throwable:\\ getMessage()};
\draw(3,-3) node (A4) {\hspace*{-0.25cm}com.stericson.RootShell.execution.Shell:\hspace*{-0.25cm}\\ add()};
\draw[->] (A0) -- (A1);
\draw[->] (A0) -- (A2);
\draw[->] (A0) -- (A3);
\draw[->] (A2) -- (A4);
\end{tikzpicture}
}
\caption{Call graph of the API calls in the try/catch block of \figurename~\ref{lst:java}. 
(Return types and parameters are omitted to ease presentation).} \label{fig:calls}
\vspace{-0.35cm}
\end{center}
\end{figure}

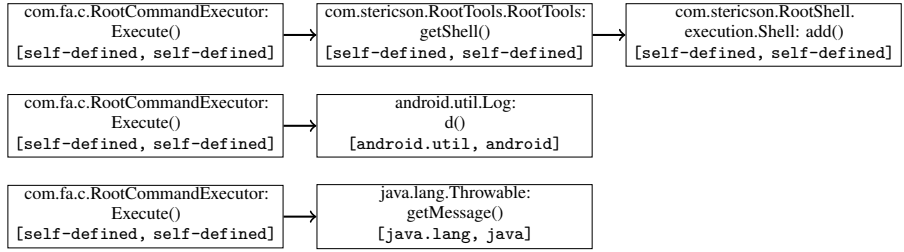
\begin{figure*}[t!]
\begin{center}
\resizebox{0.65\textwidth}{!}{\scriptsize
\begin{tikzpicture}[every node/.style={draw,rectangle},text height=0.1cm,text width=3.62cm,align=center]
\draw(0,0) node (A0) {com.fa.c.RootCommandExecutor: Execute()\\{\tt [self-defined, self-defined]}};
\draw(0,-1.25) node (A1) {com.fa.c.RootCommandExecutor:\\Execute()\\{\tt [self-defined, self-defined]}};
\draw(0,-2.5) node (A6) {com.fa.c.RootCommandExecutor:\\Execute()\\{\tt [self-defined, self-defined]}};
\draw(4.25,0) node (A2) {com.stericson.RootTools.RootTools: getShell()\\{\tt [self-defined, self-defined]}};
\draw(8.5,0) node (A3) {com.stericson.RootShell.\\execution.Shell: add()\\{\tt [self-defined, self-defined]}};
\draw(4.25,-1.25) node (A4) {android.util.Log:\\ d()\\{\tt [android.util, android]}};
\draw(4.25,-2.5) node (A5) {java.lang.Throwable:\\getMessage()\\{\tt [java.lang, java]}};
\draw[->,thick] (A0) -- (A2.west);
\draw[->,thick] (A2) -- (A3.west);
\draw[->,thick] (A1) -- (A4.west);
\draw[->,thick] (A6) -- (A5.west);

\end{tikzpicture}
}
\caption{Sequence of API calls extracted from the call graphs in
\figurename~\ref{fig:calls}, with the corresponding package/family abstraction in square brackets.} \label{fig:sequence}
\vspace{-0.5cm}
\end{center}
\end{figure*}

When operating in package mode, we abstract an API call to its package name  %
using the list of Android packages\footnote{\url{https://developer.android.com/reference/packages.html}}, which as of API level 24 (the current version as of September 2016) includes 243 packages, as well as 95 from the Google API.\footnote{\url{https://developers.google.com/android/reference/packages}} Moreover, we abstract developer-defined
packages (e.g., com.stericson.roottools) as well as obfuscated ones (e.g.
com.fa.a.b.d), respectively, as {\tt self-defined} and {\tt obfuscated}. %
Note that we label an API call's package as obfuscated if we cannot tell what its class implements, extends, or inherits, due to identifier mangling~\cite{schulz2012code}. 
When operating in family mode, we abstract to nine possible families, i.e.,  {\tt android}, {\tt google}, {\tt java}, {\tt javax}, {\tt xml}, {\tt apache}, {\tt junit}, {\tt json}, {\tt dom}, 
which correspond to the {android.*}, {com.google.*}, {java.*}, {javax.*}, {org.xml.*}, {org.apache.*},
junit.*, org.json, and org.w3c.dom.* packages.
Again, API calls from developer-defined and obfuscated packages are abstracted to families labeled as {\tt self-defined} and {\tt obfuscated}, respectively. %
Overall, there are 340 (243$+$95$+$2) possible packages and 11 (9$+$2) families. %
In \figurename~\ref{fig:sequence}, we show the sequence of API calls obtained from the call graph in \figurename~\ref{fig:calls}.
We also report, in square brackets, the family and the package to
which the call is abstracted.

\subsection{Markov-chain Based Modeling} \label{sec:MaCha}
Next, \approach builds feature vectors, used for classification, based on the
Markov chains representing the sequences of extracted API calls for an app. Before discussing this in detail, 
we review the basic concepts of Markov chains.

Markov chains are memoryless models where the probability
of transitioning from a state to another only depends on the current state~\cite{MarCha}. 
Markov chains are often represented as a set of nodes, each corresponding to a
different state, and a set of edges connecting one node to another labeled with
the probability of that transition. The sum of all probabilities associated to
all edges from any node (including, if present, an edge going back to the node
itself) is exactly 1.%
The set of possible states of the Markov chain is denoted as $\mathcal{S}$. If $S_{j}$ and $S_{k}$ are two connected states, $P_{jk}$ denotes the probability of transition from $S_{j}$ to $S_{k}$. $P_{jk}$ is given by the number of 
occurrences ($O_{jk}$) of state $S_{k}$ after state $S_{j}$, divided by $O_{ji}$ for all states $i$ in the chain, i.e.,
$P_{jk}=\frac{O_{jk}}{\sum_{i \in \mathcal{S}} O_{ji}}$.

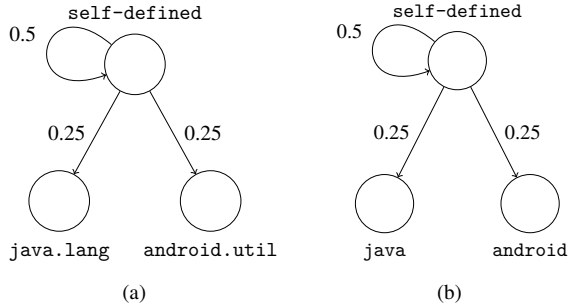
\begin{figure}[t!]
\begin{center}
\subfigure[]{
\resizebox{.215\textwidth}{!}{
\begin{tikzpicture}
\tikzstyle{level 1}=[level distance=1.75cm, sibling distance=2.5cm]
\tikzset{
myroot/.style={label=above:{\tt\strut#1},align=center,anchor=north,draw,circle,minimum size=1cm},
myleaf/.style={label=below:{\tt\strut#1},align=center,anchor=north,draw,circle,minimum size=1cm},
every loop/.style={max distance=200mm,in=-160,out=-220}}
 \node[myroot=self-defined] (q0) {}
    child {node[myleaf=java.lang] (q1) {} edge from parent[draw=none]} 
    child {node[myleaf=android.util] (q2) {}  edge from parent[draw=none]}; 
\draw[->] (q0) edge[loop above] () node [midway] {\hspace*{-3.7cm}{0.5}};    
\draw[->] (q0) -- (q1) node [midway=10pt] {\hspace{-1cm}0.25};
\draw[->] (q0) -- (q2) node [midway=10pt] {\hspace{1cm}0.25};
\end{tikzpicture}
}}
\hspace{0.2cm}
\subfigure[]{
\resizebox{.178\textwidth}{!}{
\begin{tikzpicture}
\tikzstyle{level 1}=[level distance=1.95cm, sibling distance=2.5cm]
\tikzset{
myroot/.style={label=above:{\tt\strut#1},align=center,anchor=north,draw,circle,minimum size=1cm},
myleaf/.style={label=below:{\tt\strut#1},align=center,anchor=north,draw,circle,minimum size=1cm},
every loop/.style={max distance=200mm,in=-160,out=-220}}
 \node[myroot=self-defined] (q0) {} 
    child {node[myleaf=java] (q1) {}edge from parent[draw=none]} 
    child {node[myleaf=android] (q2) {}edge from parent[draw=none]}; 
\draw[->] (q0) edge[loop above] () node [midway] {\hspace*{-3.7cm}{0.5}};    
\draw[->] (q0) -- (q1) node [midway=10pt] {\hspace{-1cm}0.25};
\draw[->] (q0) -- (q2) node [midway=10pt] {\hspace{1cm}0.25};
\end{tikzpicture}
}}
\end{center}
\vspace{-0.35cm}
\caption{Markov chains originating from the call sequence example in Section~\ref{sec:sequence} when using packages (a) or families (b).}
\label{fig:package}
\vspace{-0.3cm}
\end{figure}

\descr{Building the model.} \approach uses Markov chains to model app behavior, by evaluating every transition between calls. More specifically, for each app, \approach 
takes as input the sequence of
abstracted API calls of that app -- i.e., packages or families, depending on the selected mode of operation -- 
and builds a Markov chain where each package/family is a
state and the transitions represent the probability of moving from one state to another. For each Markov chain, state $S_0$ is the entry point from which other calls are made in a sequence.
As an example, \figurename~\ref{fig:package} illustrates the two Markov chains built using packages and families, respectively, 
from the sequences reported in \figurename~\ref{fig:sequence}.%

We argue that considering single transitions is more robust against attempts to evade detection
by inserting useless API calls in order to deceive signature-based
systems (see Section~\ref{sec:related}). %
In fact, \approach considers all possible calls -- i.e., all the branches originating from a node -- in the Markov chain, 
so adding calls would not significantly change the probabilities of transitions between nodes (specifically, families or packages, depending on the operational mode) for each app. %

\descr{Feature Extraction.} 
Next, we use the probabilities of transitioning from one state (abstracted call) to another in the Markov chain
as the feature vector of each app. States that are not present in a chain are represented as 0 in the feature vector. Also note that 
the vector derived from the Markov chain depends on the operational mode of \approach. 
With families, there are 11 possible states, thus 121 possible transitions in each chain, while, when abstracting to packages, there are 340 states and 115,600 possible transitions.

We also apply Principal Component Analysis (PCA)~\cite{PCA}, which performs feature selection by transforming the feature space into a new space made of components that are a linear combination of the original features. The first components contain as much variance (i.e., amount of information) as possible. The variance %
is given as percentage of the total amount of information of the original
feature space. We apply PCA to the feature set in order to select the principal
components, as PCA  transforms the feature space into a smaller one where the
variance is represented with as few components as possible, thus considerably
reducing computation/memory complexity. 
Furthermore, the use of PCA could also improve the accuracy of the classification, by taking misleading features out of the feature space,
i.e., those that make the classifier perform worse.

\subsection{Classification}\label{sec:classification}
The last step is to perform classification, i.e., labeling apps as either benign or
malware. To this end, we test \approach using different classification algorithms: Random
Forests~\cite{RandomForest}, 1-Nearest Neighbor (1-NN)~\cite{KNN}, 3-Nearest
Neighbor (3-NN)~\cite{KNN}, and Support Vector Machines (SVM)~\cite{SVM}. Each model is trained using the feature vector obtained from the apps in a training sample. Results are presented and discussed in Section \ref{sec:evaluation}, and have been validated by using 10-fold cross validation.

Also note that, due to the different number of features used in family/package modes, we use two
distinct configurations for the Random Forests algorithm. Specifically, when abstracting to
families, we use 51 trees with maximum depth 8, while, with packages, we use 101 trees of maximum depth
64. To tune Random Forests we followed the methodology applied in~\cite{RandTune}.

\begin{table*}[t]
\centering
\small
\resizebox{1.25\columnwidth}{!}{
\begin{tabular}{|l|l|ll|rrr|}
\multicolumn{7}{c}{}\\
\hline
{\bf Category} & {\bf Name} & \multicolumn{2}{l|}{\bf Date Range}  & {\bf \#Samples} & {\bf \#Samples}  & {\bf \#Samples}\\
& & & & & {\bf (API Calls)} & {\bf (Call Graph)}\\
\hline
\multirow{2}{*}{\em Benign} & {\tt oldbenign} & Apr 2013\hspace*{-0.15cm}&-- Nov 2013  & 5,879 & 5,837 & 5,572 \\
 & {\tt newbenign} & Mar 2016\hspace*{-0.15cm} &-- Mar 2016 & 2,568 & 2,565 & 2,465\\
 \hline
 \multicolumn{4}{|r}{\em Total Benign:} & \multicolumn{1}{r}{\em 8,447} & \multicolumn{1}{r}{\em 8,402} & \multicolumn{1}{r|}{\em 8,037}\\[1ex]
\hline
\multirow{5}{*}{\em Malware} & {\tt drebin} & Oct 2010\hspace*{-0.15cm} &-- Aug 2012 & 5,560 & 5,546 & 5,512\\
 & {\tt 2013} &  Jan 2013\hspace*{-0.15cm} &-- Jun 2013 & 6,228 & 6,146 & 6,091\\
& {\tt 2014} & Jun 2013\hspace*{-0.15cm} &-- Mar 2014 &  15,417 & 14,866 & 13,804\\
 & {\tt 2015} & Jan 2015\hspace*{-0.15cm} &-- Jun 2015  & 5,314 & 5,161 & 4,451\\
 & {\tt 2016} & Jan 2016\hspace*{-0.15cm} &-- May 2016 & 2,974 & 2,802 & 2,555 \\
 \hline
  \multicolumn{4}{|r}{\em Total Malware:} & \multicolumn{1}{r}{\em 35,493} & \multicolumn{1}{r}{\em 34,521} & \multicolumn{1}{r|}{\em 32,413}\\[0.5ex]
  \hline
\end{tabular}
}
\vspace{-0.1cm}
\caption{Overview of the datasets used in our experiments.} %
\label{table:dataset}
\vspace{-0.5cm}
\end{table*}

\section{Datasets}
\label{sec:data}

In this section, we introduce the datasets used in the evaluation of \approach (presented later in Section~\ref{sec:evaluation}),
which include 43,940 apk files -- 8,447 benign and 35,493 malware samples.
We include a mix of older and newer apps, ranging from October 2010 to May 2016, 
as we aim to verify that \approach is robust to changes in Android malware samples as well as APIs.
To the best of our knowledge, we are leveraging the largest dataset of malware samples ever used in a
research paper on Android malware detection.

\descr{Benign Samples.} Our benign datasets consist of two sets of samples: (1) one, which we denote as {\tt oldbenign}, includes 5,879 apps collected by PlayDrone~\cite{viennot2014measurement} between April and November 2013, and published on the Internet Archive\footnote{\url{https://archive.org/details/playdrone-apk-e8}} on August 7, 2014; and (2) another, {\tt newbenign}, 
obtained by downloading the top 100 apps in each of the 29 categories on the Google Play store\footnote{\url{https://play.google.com/store}} as of March 7, 2016, using the googleplay-api tool.\footnote{\url{https://github.com/egirault/googleplay-api}} Due to errors encountered while downloading some apps, we have actually obtained 2,843 out of 2,900 apps. Note that 275 of these belong to more than one category, therefore, the \texttt{newbenign} dataset ultimately includes 2,568 unique apps.

\descr{Android Malware Samples.} The set of malware samples includes apps that were used to test
{\sc Drebin}~\cite{arp2014drebin}, dating back to October 2010 -- August 2012
(5,560), which we denote as {\tt drebin}, as well as more recent ones that have
been uploaded on the VirusShare\footnote{\url{https://virusshare.com/}} site over the years. Specifically, we gather from VirusShare, respectively, 6,228, 15,417, 5,314, and 2,974 samples from {\tt 2013}, {\tt 2014}, {\tt 2015}, and {\tt 2016}. We consider each of these datasets separately for our analysis.

\descr{API Calls and Call Graphs.} For each app in our datasets, we extract the list of API calls, using Androguard\footnote{\url{https://github.com/androguard/androguard}}, since, as explained in Section~\ref{sec:compare}, these constitute the features used by \droid~\cite{Aafer2013DroidAPIMiner}, against which we compare our system. Due to Androguard failing to decompress some of the apks, bad CRC-32 redundancy checks, and errors during unpacking, we are not able to extract the API calls for all the samples, but
only for 40,923 (8,402 benign, 34,521 malware) out of the 43,940 apps (8,447 benign, 35,493 malware) in our datasets.
Also, to extract the call graph of each apk, we use Soot. 
Note that for some of the larger apks, Soot requires a non-negligible amount of memory to extract the call graph, so 
we allocate 16GB of RAM to the Java VM heap space. %
We find that for 
2,472 (364 benign + 2,108 malware) samples, Soot is not able to complete the
extraction due to it failing to apply the {\tt jb} phase as well as reporting an
error in opening some zip files (i.e., the apk).  The {\tt jb} phase is used by Soot to transform Java bytecode into jimple intermediate representation (the primary IR of Soot) for optimization purposes. %
Therefore, we exclude these apps in our evaluation and discuss this limitation further in Section~\ref{sec:limits}. %
In Table~\ref{table:dataset}, we provide a summary of our seven datasets, reporting the total number of samples per dataset,
as well as those for which we are able to extract the API calls (second-to-last column) and the call graphs (last column). 

\begin{figure}[t!]
\centering
\includegraphics[width=0.405\textwidth]{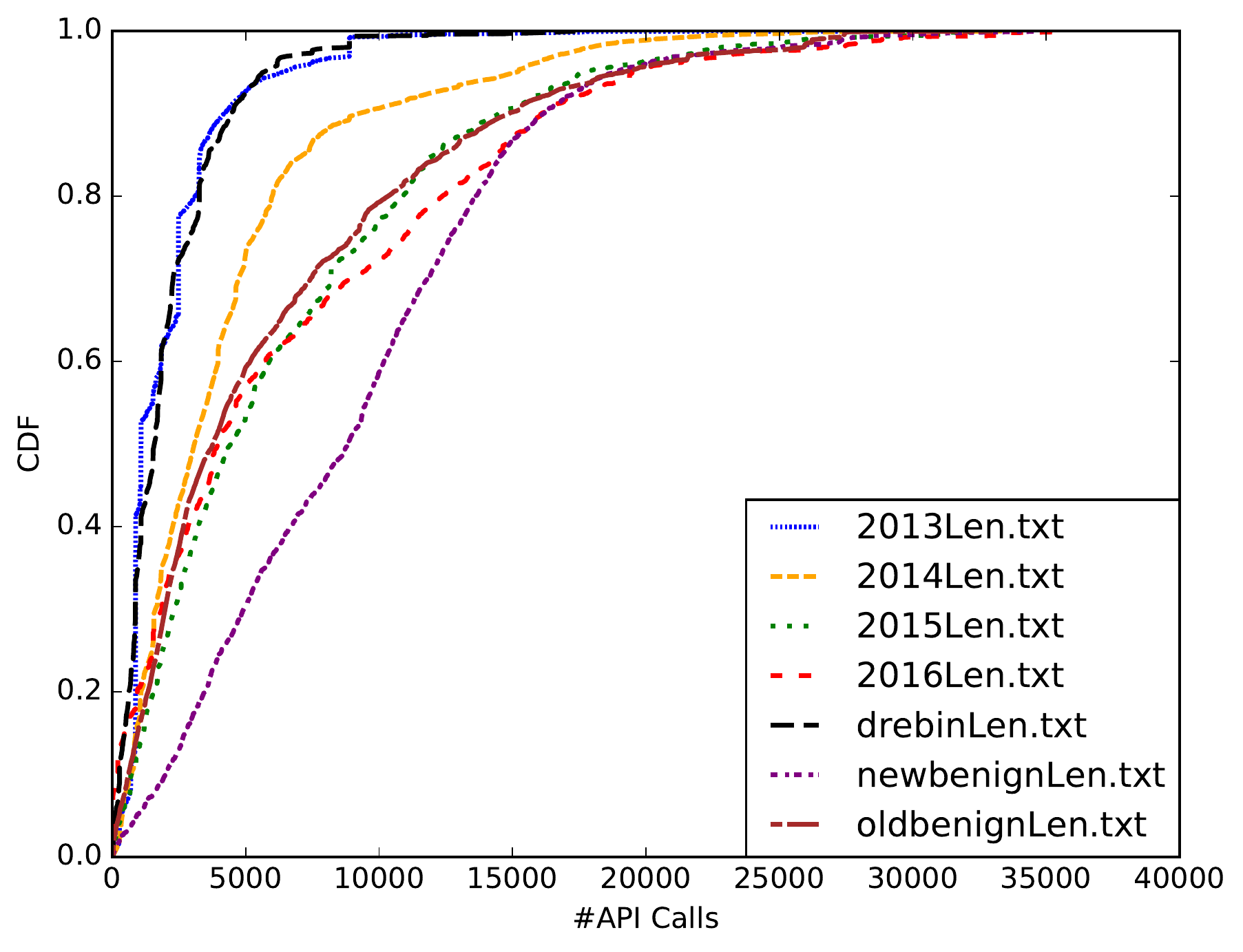}
\vspace{-0.2cm}
\caption{CDF of the number of API calls in different apps in each dataset.}
\label{fig:NumCalls}
\vspace{-0.3cm}
\end{figure}

\begin{figure}[t]
\centering
\subfigure[\label{fig:FamAndro}{\tt android}]
{\includegraphics[width=0.405\textwidth]{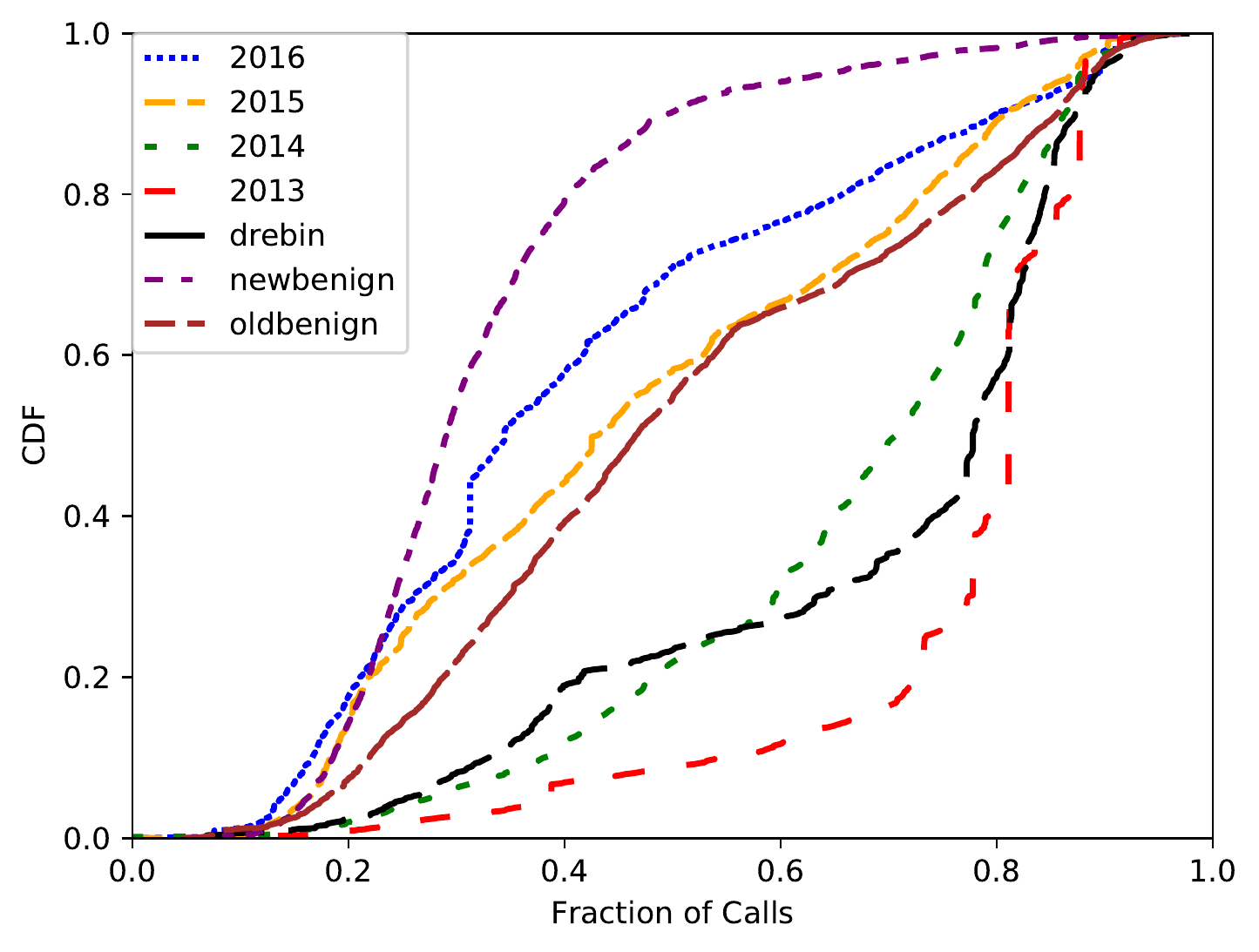}}
\\[-0.75ex]
\subfigure[\label{fig:FamGoogle}{\tt google}]
{\includegraphics[width=0.405\textwidth]{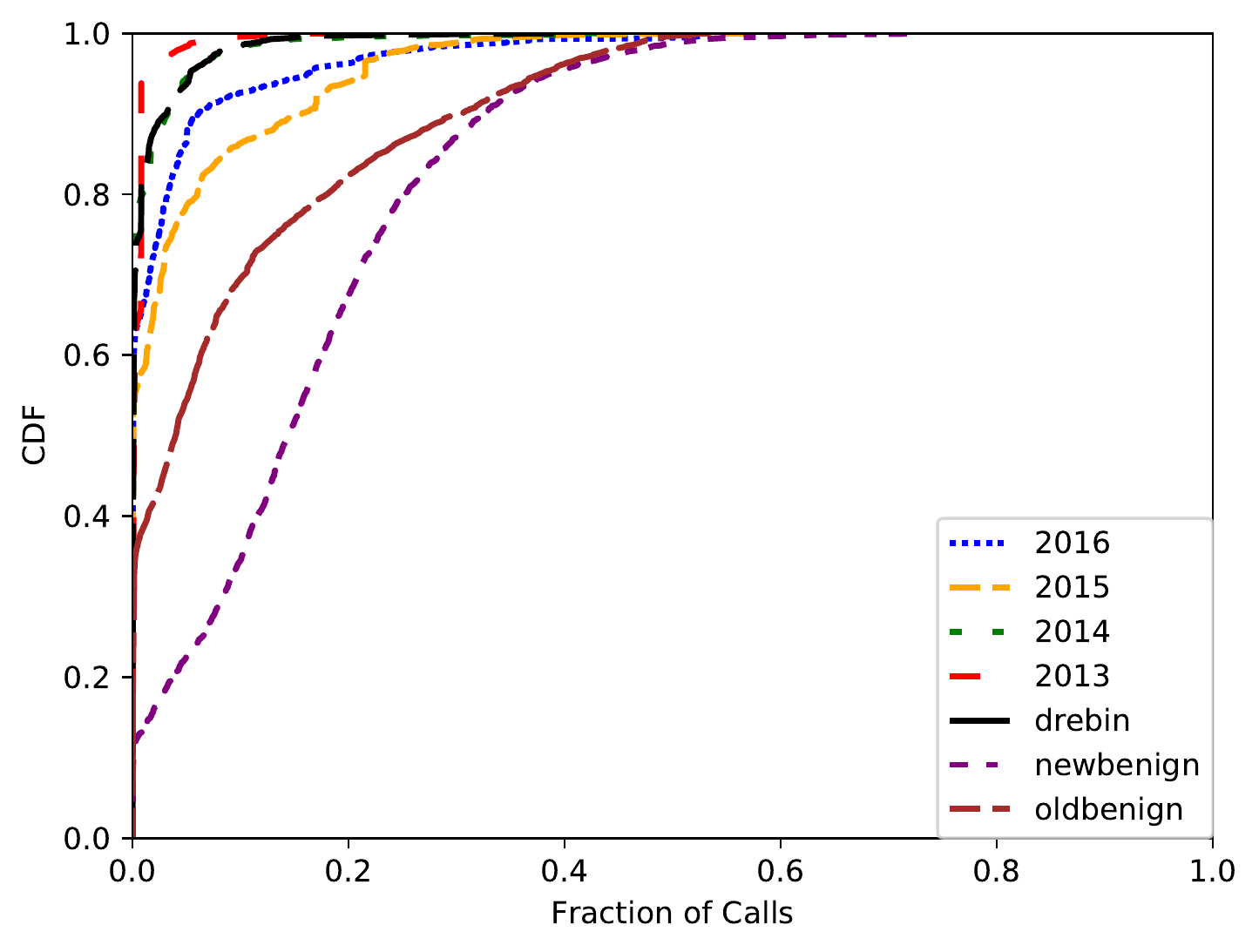}}
\vspace{-0.25cm}
\caption{CDFs of the percentage of {\tt android} and {\tt google} family calls in different apps in each dataset.}
\vspace{-0.3cm}
\end{figure}

\begin{figure}[t]
\centering
\subfigure[\label{fig:PCAbenign}{benign}]
{\includegraphics[width=0.405\textwidth]{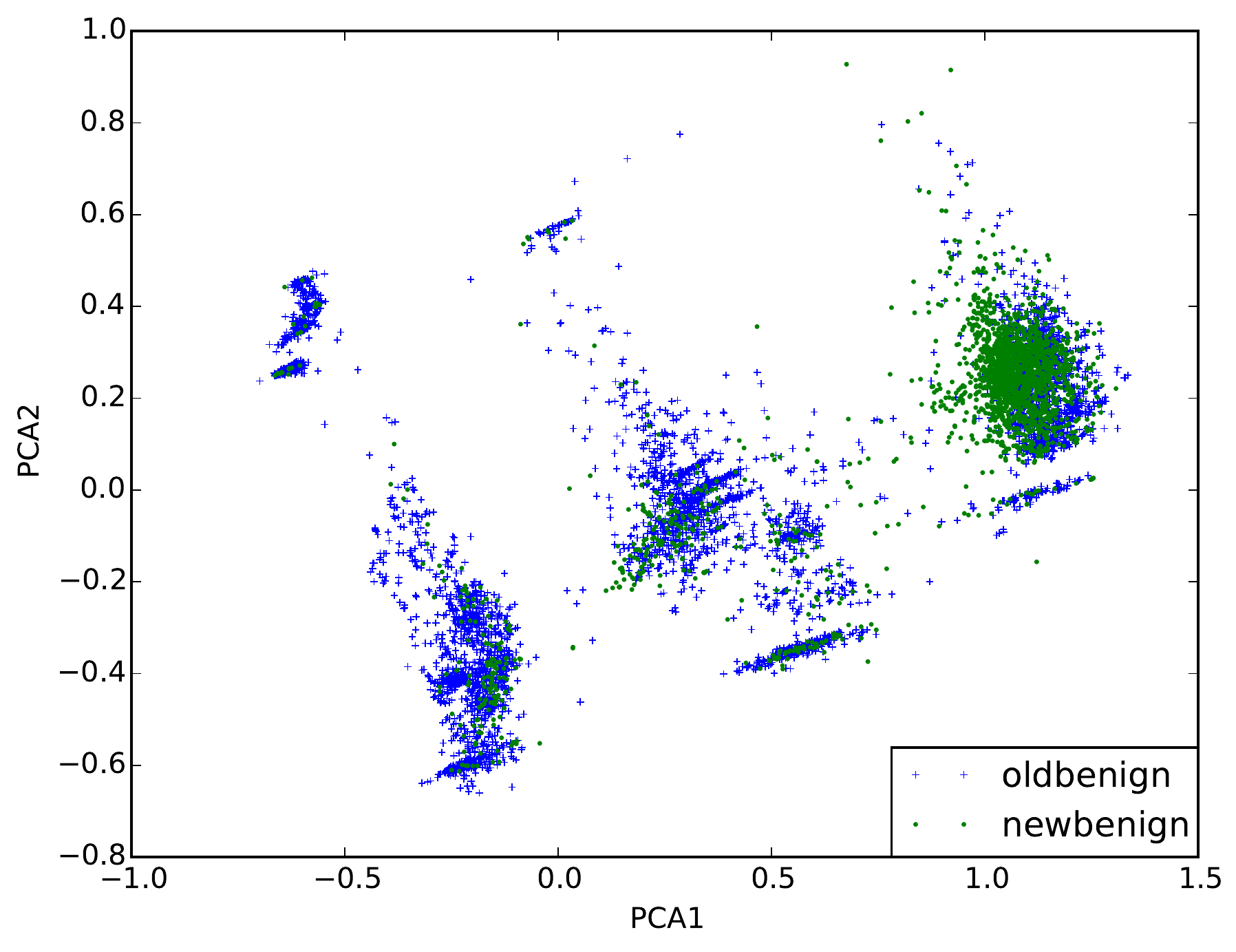}}
\\[-0.75ex]
\subfigure[\label{fig:PCAmalicious}{malware}]
{\includegraphics[width=0.405\textwidth]{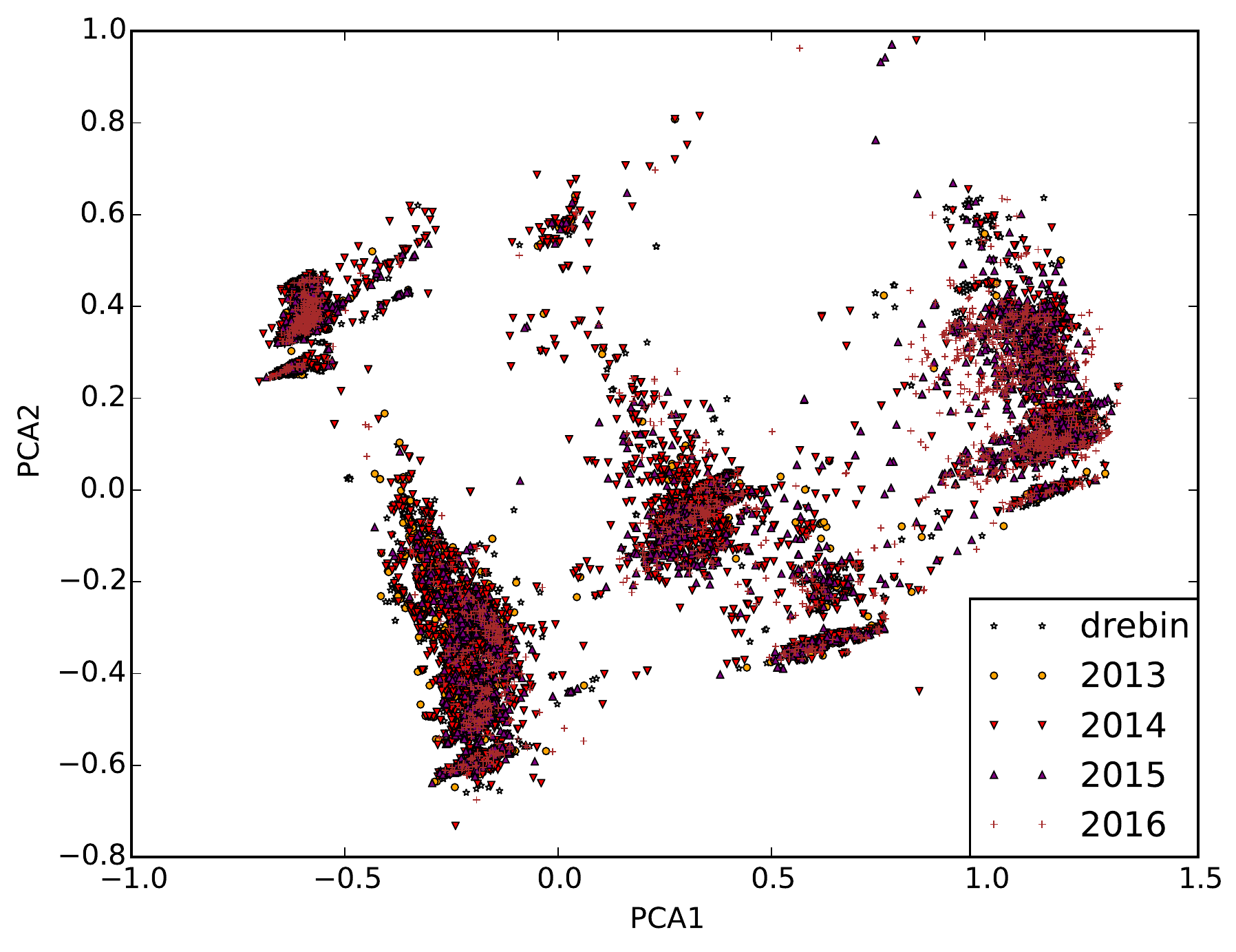}}
\vspace{-0.25cm}
\caption{Positions of benign vs malware samples in the feature space of the first two components of the PCA (family mode).}
\vspace{-0.3cm}
\end{figure}

\descr{Characterization of the Datasets.} Aiming to shed light on the evolution of API calls in Android apps,
we also performed some measurements over our datasets. %
In \figurename~\ref{fig:NumCalls}, we plot the Cumulative Distribution Function
(CDF) of the number of unique API calls in the apps in different datasets, highlighting that newer apps, both
benign and malicious, are using more API calls overall than older apps. This
indicates that as time goes by, Android apps become more complex.
When looking at the fraction of API calls belonging to specific families, we
discover some interesting aspects of Android apps developed in different years.
In particular, we notice that API calls to the {\tt android} family become less
prominent as time passes (\figurename~\ref{fig:FamAndro}), both in benign and
malicious datasets, while {\tt google}
calls become more common in newer apps (\figurename~\ref{fig:FamGoogle}).

In general, we conclude that benign and malicious apps show the same
evolutionary trends over the years. Malware, however, appears to reach the same
characteristics (in terms of level of complexity and fraction of API calls from
certain families) as legitimate apps with a few years of delay.

\descr{Principal Component Analysis.} Finally, we apply PCA to select the two most important PCA components. We
plot and compare the positions of the two components  for benign (\figurename~\ref{fig:PCAbenign}) and malicious samples (Fig. \ref{fig:PCAmalicious}). 
As PCA combines the features into components, it maximizes the variance of the distribution of
samples in these components, 
thus, plotting the positions of the samples in the components shows that benign apps tend to be located in different areas of the components space, depending on the dataset, while malware samples occupy similar areas but with different densities. These differences highlight a different behavior between benign and malicious samples, and these differences should also be found by the machine learning algorithms 
used for classification.

\section{Evaluation}
\label{sec:evaluation}
We now present a detailed experimental evaluation of \approach. Using
the datasets %
summarized in Table~\ref{table:dataset},
we perform four sets of experiments: (1) we analyze the accuracy of \approach's classification 
on benign and malicious samples developed around the same time; (2) we evaluate its robustness to the evolution of malware as well as of the Android framework by using older datasets for training and newer ones for testing (and vice-versa); (3) we measure \approach's runtime performance to assess its scalability; and, finally, (4) we compare against \droid~\cite{Aafer2013DroidAPIMiner}, a malware detection system that relies on the frequency of API calls. %

\subsection{Preliminaries}
When implementing \approach in family mode, we exclude %
the {\tt json} and {\tt dom}  families
because they are almost never used across all our datasets, and {\tt junit}, which is primarily used for testing.
In package mode, to avoid mislabeling when {\tt self-defined} APIs have ``android'' in the name, we split the {\tt android} package into its  two classes, i.e., {\tt android.R} and {\tt android.Manifest}.
Therefore, in family mode, there are 8 possible states, thus 64 features, whereas, in package mode, we have 341 states and 116,281 features (cf.~Section~\ref{sec:MaCha}). %

As discussed in Section~\ref{sec:classification}, we use four different machine learning algorithms for classification -- namely, Random Forests~\cite{RandomForest}, 1-NN~\cite{KNN}, 3-NN~\cite{KNN}, and SVM~\cite{SVM}. %
Since both accuracy and speed are worse with SVM than with the other three algorithms, 
we omit results obtained with SVM. %
To assess the accuracy of the classification, we use the standard F-measure metric, i.e.: \vspace{-0.15cm}
\begingroup\makeatletter\def\f@size{9}\check@mathfonts
$$\small F=2\cdot\dfrac{\mbox{precision}\cdot\mbox{recall}}{\mbox{precision}+\mbox{recall}}$$\endgroup
where precision $=$ TP$/($TP$+$FP$)$ and 
recall $=$ TP$/($TP$+$FN$)$.
TP denotes the number of samples correctly classified as malicious, while
FP an FN indicate, respectively, the number of samples mistakenly identified as malicious and benign. 

Finally, note that all our experiments perform 10-fold cross validation using at
least one malicious and one benign dataset from Table~\ref{table:dataset}. In
other words, after merging the datasets, the resulting set is shuffled and
divided into ten equal-size random subsets. Classification is then performed ten
times using nine subsets for training and one for testing, and results are
averaged out over the ten experiments.

\begin{figure}[t]
\centering
        \includegraphics[width=0.495\textwidth]{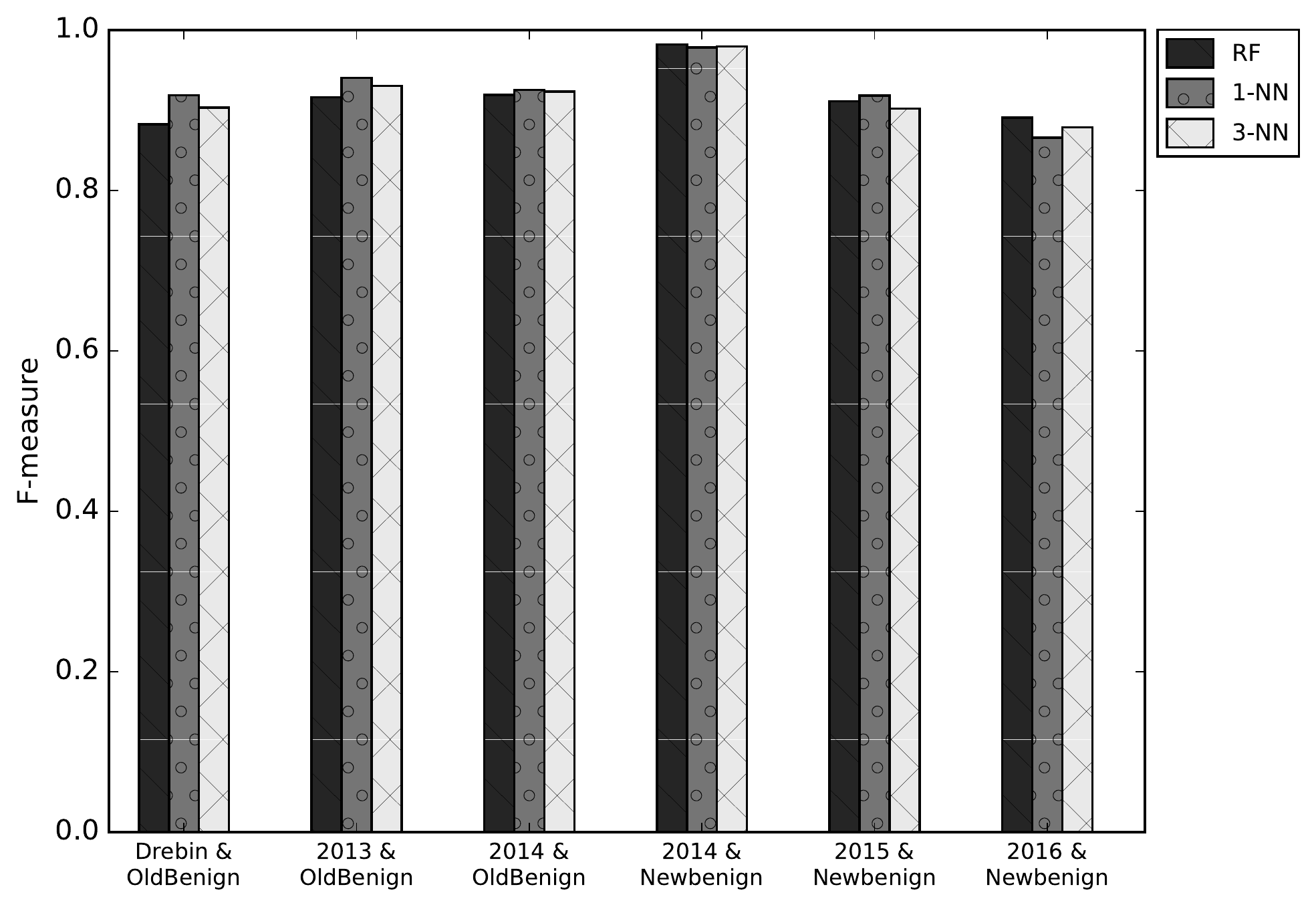} %
	\vspace{-0.15cm}
    \caption{F-measure of \approach classification with datasets from the same year (family mode).}
    \label{fig:FM10fam}
    	\vspace{-0.25cm}
\end{figure}

\subsection{Detection Performance}

\begin{table*}[t]
\centering
\small
\setlength{\tabcolsep}{3pt}
\resizebox{0.99\linewidth}{!}{
\begin{tabular}{|l|p{1cm}rr|p{0.85cm}rr|p{0.85cm}rr|p{0.85cm}rr|p{0.85cm}rr|p{0.85cm}rr|}
\hline
\multirow{ 2}{*}{\backslashbox{\bf Mode}{\bf \hspace{-0.2cm}Dataset\hspace{-0.2cm}}}  & \multicolumn{18}{c|}{[Precision, Recall, {\bf $\textbf{F}$-measure}]}   \\
\cline{2-19}
 & \multicolumn{3}{r|}{\texttt{drebin} \& {\tt oldbenign}} & \multicolumn{3}{l|}{\texttt{2013} \& {\tt oldbenign}} & \multicolumn{3}{l|}{\texttt{2014} \& {\tt oldbenign}} & \multicolumn{3}{l|}{\texttt{2014} \& {\tt newbenign}} & \multicolumn{3}{l|}{\texttt{2015} \& {\tt newbenign}} & \multicolumn{3}{l|}{\texttt{2016} \& {\tt newbenign}}\\
\hline
Family &  0.82 & 0.95 & {\bf 0.88} & 0.91 & 0.93 & {\bf 0.92} & 0.88 & 0.96 & {\bf 0.92} & 0.97 & 0.99 & {\bf 0.98} & 0.89 & 0.93 & {\bf 0.91} & 0.87 & 0.91 & {\bf 0.89} \\
\hline
Package & 0.95 & 0.97 &  {\bf 0.96} & 0.98 &  0.95  &  {\bf 0.97} & 0.93 &  0.97 &  {\bf 0.95} & 0.98  & 1.00 &  {\bf 0.99} & 0.93 &  0.98 &  {\bf 0.95} & 0.92 &  0.92 &  {\bf 0.92} \\
\hline
Family (PCA) & 0.84 &  0.92 &  {\bf 0.88} & 0.93 &  0.90 &  {\bf 0.92} & 0.87 &  0.94 &  {\bf 0.90} & 0.96 &  0.99 &  {\bf 0.97} & 0.87 &  0.93 & {\bf 0.90} & 0.86 &  0.88 &  {\bf 0.87} \\
\hline
Package (PCA) & 0.94 &  0.95 &  {\bf 0.94} & 0.97 &  0.95 &  {\bf 0.96} & 0.92 & 0.96 &  {\bf 0.94} & 0.97 & 1.00 &  {\bf 0.99} & 0.91&  0.97&  {\bf 0.94} & 0.88 & 0.89&  {\bf 0.89} \\
\hline
\end{tabular}
}
\caption{F-measure, precision, and recall obtained by \approach, using Random Forests, on various dataset combinations with different modes of operation, with and without PCA.}
\label{table:overallresults}
\vspace{-0.4cm}
\end{table*}

We start our evaluation by measuring how well \approach detects malware by training and testing using samples
that are developed around the same time. To this end, we perform 10-fold cross validations on the
combined dataset composed of a benign set and a malicious one. Table~\ref{table:overallresults} provides an overview of the detection results achieved by \approach
on each combined dataset, in the two modes of operation, both with PCA features and without. The reported F-measure, precision, and recall scores are the ones obtained with Random Forest, which generally performs better than 1-NN and 3-NN. 

\descr{Family mode.} In \figurename~\ref{fig:FM10fam}, we report the F-measure
when operating in family mode for Random Forests, 1-NN and 3-NN. The F-measure is always at least 88\% with Random Forests, and, when tested on the {\tt 2014} (malicious) dataset, it reaches 98\%. 
With some datasets, \approach performs slightly better than with others. For instance, with the {\tt 2014} malware dataset, we obtain
 an F-measure of 92\% when using the {\tt oldbenign} dataset and 98\%  with {\tt newbenign}. In general, lower F-measures are due to increased false positives since recall is always above 91\%, while precision might be lower, also due to the fact that malware datasets are larger than the benign sets. 
We believe that this follows the evolutionary trend discussed in
Section~\ref{sec:data}: %
while both benign and malicious apps become more complex as time passes, when a
new benign app is developed, it is still possible to use old classes or re-use
code from previous versions and this might cause them to be more similar to old
malware samples. This would result in false positives by \approach. %
In general, \approach performs better when the different characteristics of malicious and benign training and test sets are more 
predominant, which corresponds to datasets occupying different positions of the feature space.%

\begin{figure}[t]
\centering
        \includegraphics[width=0.495\textwidth]{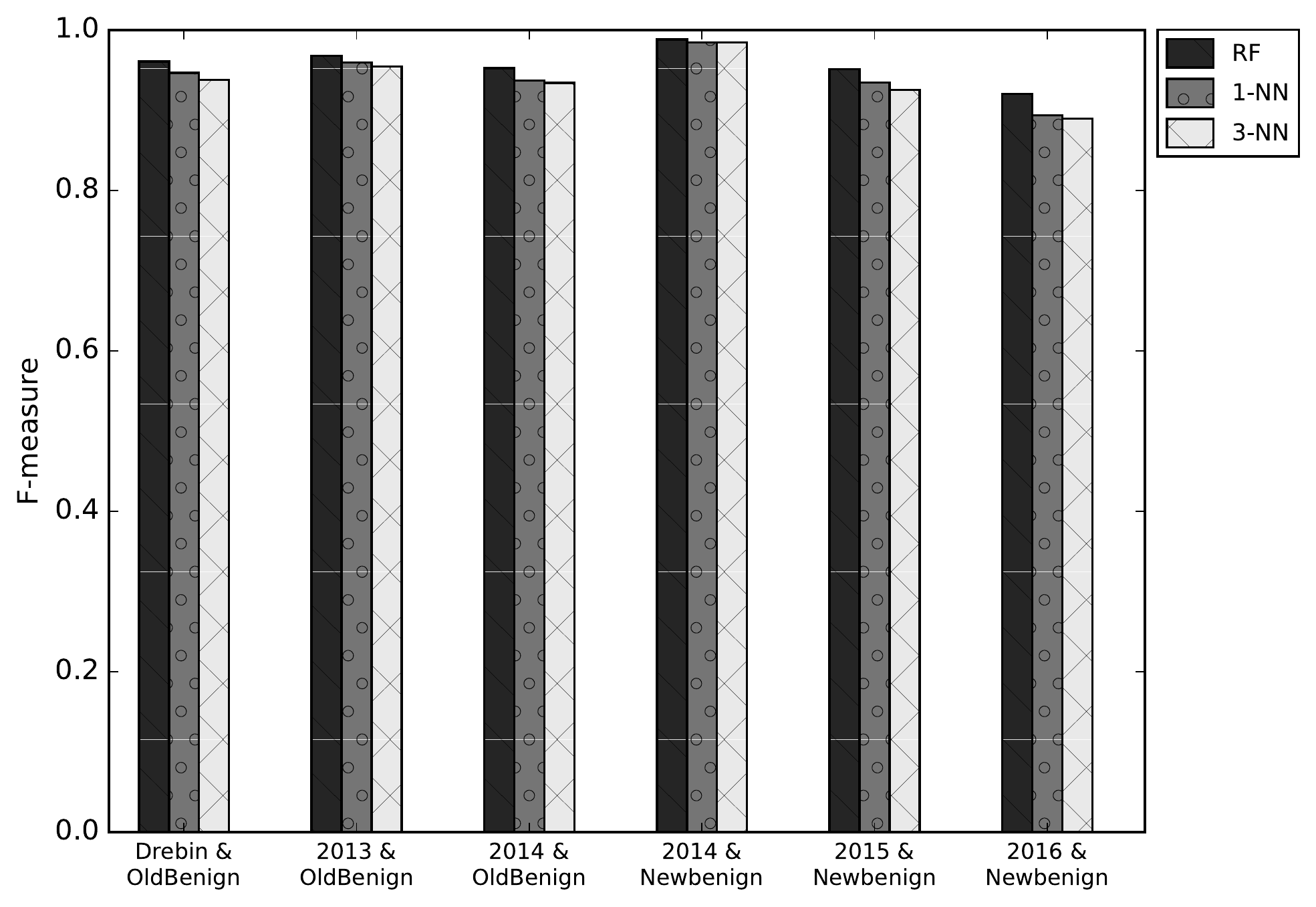} %
	\vspace{-0.15cm}
    \caption{F-measure of \approach classification with datasets from the same year (package mode).}
    \label{fig:FM10pac}
    	\vspace{-0.25cm}
\end{figure}

\descr{Package mode.} When \approach runs in package mode, the classification performance improves, %
ranging from 92\% F-measure with {\tt 2016} and {\tt newbenign} to 99\% with {\tt 2014} and {\tt newbenign}, using Random Forests. \figurename~\ref{fig:FM10pac} reports the F-measure of the 10-fold cross validation experiments using Random Forests, 1-NN, and 3-NN (in package mode). The former generally provide better results also in this case.

With some datasets, the difference in performance between the two modes of
operation is more noticeable: with {\tt drebin} and {\tt
oldbenign}, and using Random Forests, we get 96\% F-measure in package mode compared to 88\% in family mode.
These differences are caused by a lower number of false positives in package
mode. Recall remains high, resulting in a more balanced system overall.
In general, abstracting to packages rather than families provides better results
as the increased granularity enables identifying more differences between benign
and malicious apps. On the other hand, however, this likely reduces the
efficiency of the system, as many of the states deriving from the abstraction
are used a only few times. The differences in time performance between the two modes are analyzed in details in Section~\ref{sub:timeeval}.

\descr{Using PCA.} As discussed in Section~\ref{sec:MaCha}, %
PCA transforms large feature spaces into
smaller ones, thus it can be useful to significantly reduce computation and, above all, memory complexities of the classification task. 
When operating in package mode, PCA is particularly beneficial, since
\approach originally has to operate over 116,281 features. Therefore, we compare results obtained using PCA by fixing
the number of components to 10 and checking the quantity of variance included in them. 
In package mode, we observe that only 67\% of the variance is taken into account by the 10 most important PCA components,
whereas, in family mode, at least 91\% of the variance is included by the 10 PCA Components.

As shown in Table~\ref{table:overallresults}, the F-measure obtained using Random
Forests and the PCA components sets derived from the family and package features
is only slightly lower (up to 3\%) than using the full feature set. We note that
lower F-measures are caused by a uniform decrease in both precision and recall.

\subsection{Detection Over Time}
As Android evolves over the years, so do the characteristics 
of both benign and malicious apps.  Such evolution must be taken into account when evaluating Android malware detection systems, since their accuracy might significantly be affected as newer APIs are released and/or as malicious developers modify their strategies in order to avoid detection. Evaluating this aspect constitutes one of our research questions, and one of the reasons why our datasets span across multiple years (2010--2016).

As discussed in Section~\ref{sec:extraction}, \approach relies on the sequence of API calls extracted from the
call graphs and abstracted at either the package or the family level. Therefore, it is less susceptible to changes in the Android API
than other classification systems such as \droid~\cite{Aafer2013DroidAPIMiner} 
and {\sc Drebin}~\cite{arp2014drebin}. Since these rely on the
use, or the frequency, of certain API calls to classify malware vs benign samples, they need to be
retrained following new API releases. On the contrary, retraining is not needed as often with \approach, since families and packages represent
more abstract functionalities that change less over time.
Consider, for instance, the {\tt android.os.health} package: released with API level 24, it contains a set of
classes helping developers track and monitor system
resources.\footnote{\url{https://developer.android.com/reference/android/os/health/package-summary.html}} Classification systems built before this release -- as in the case of \droid~\cite{Aafer2013DroidAPIMiner} (released in 2013, when Android API was up to level 20) --
need to be retrained if this package is more frequently used by malicious apps than benign apps, while \approach only needs to add a new state to its Markov chain when operating in package mode, while no additional state is required when operating in family mode.  

\begin{figure}[t]
\centering
        \includegraphics[width=0.405\textwidth]{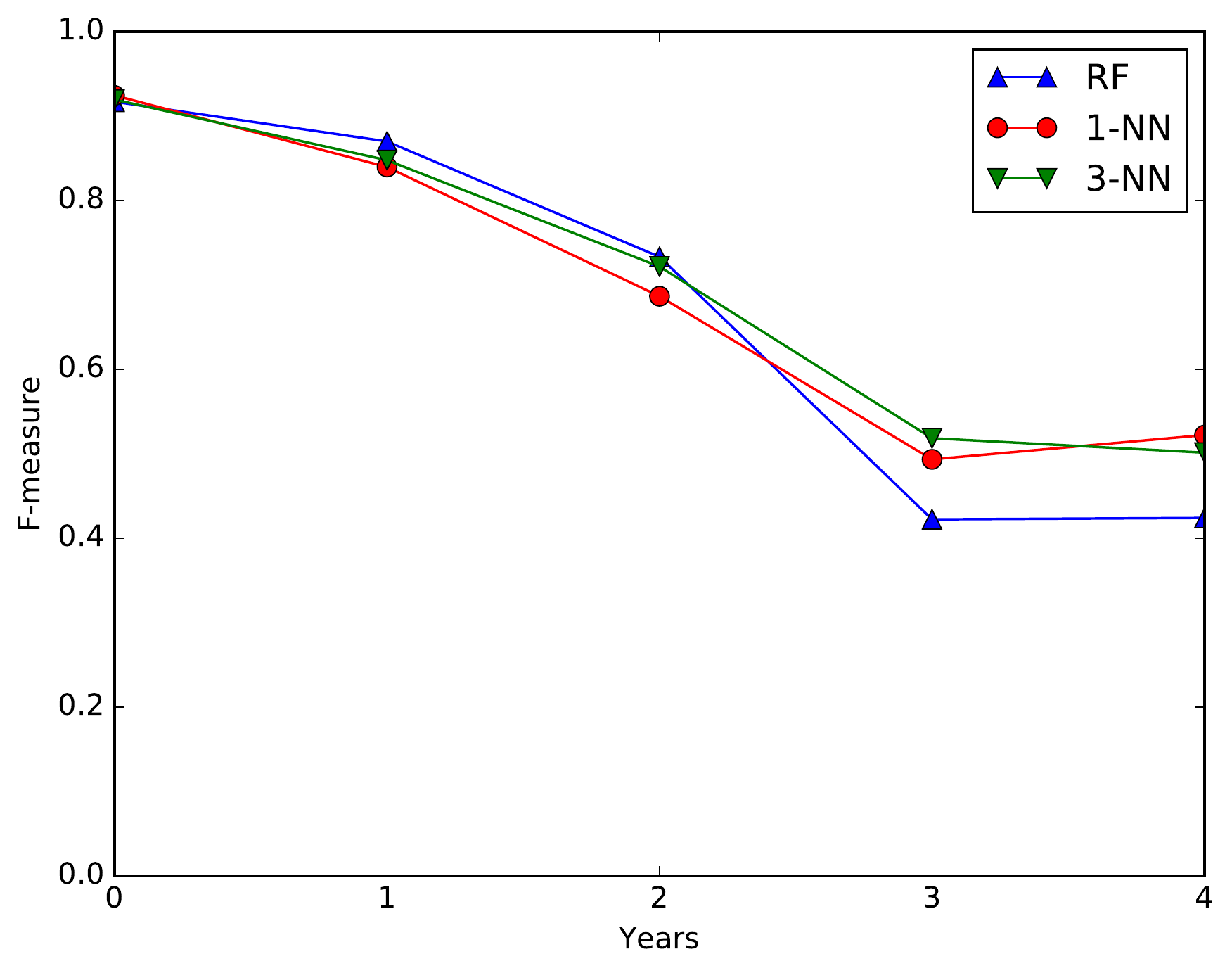} %
        \vspace{-0.2cm}
	\caption{F-measure of \approach classification using older samples for training and newer for
	testing (family mode).}
    \label{fig:FMFutfam}
        \vspace{-0.2cm}
\end{figure}

\begin{figure}[t]
\centering
        \includegraphics[width=0.405\textwidth]{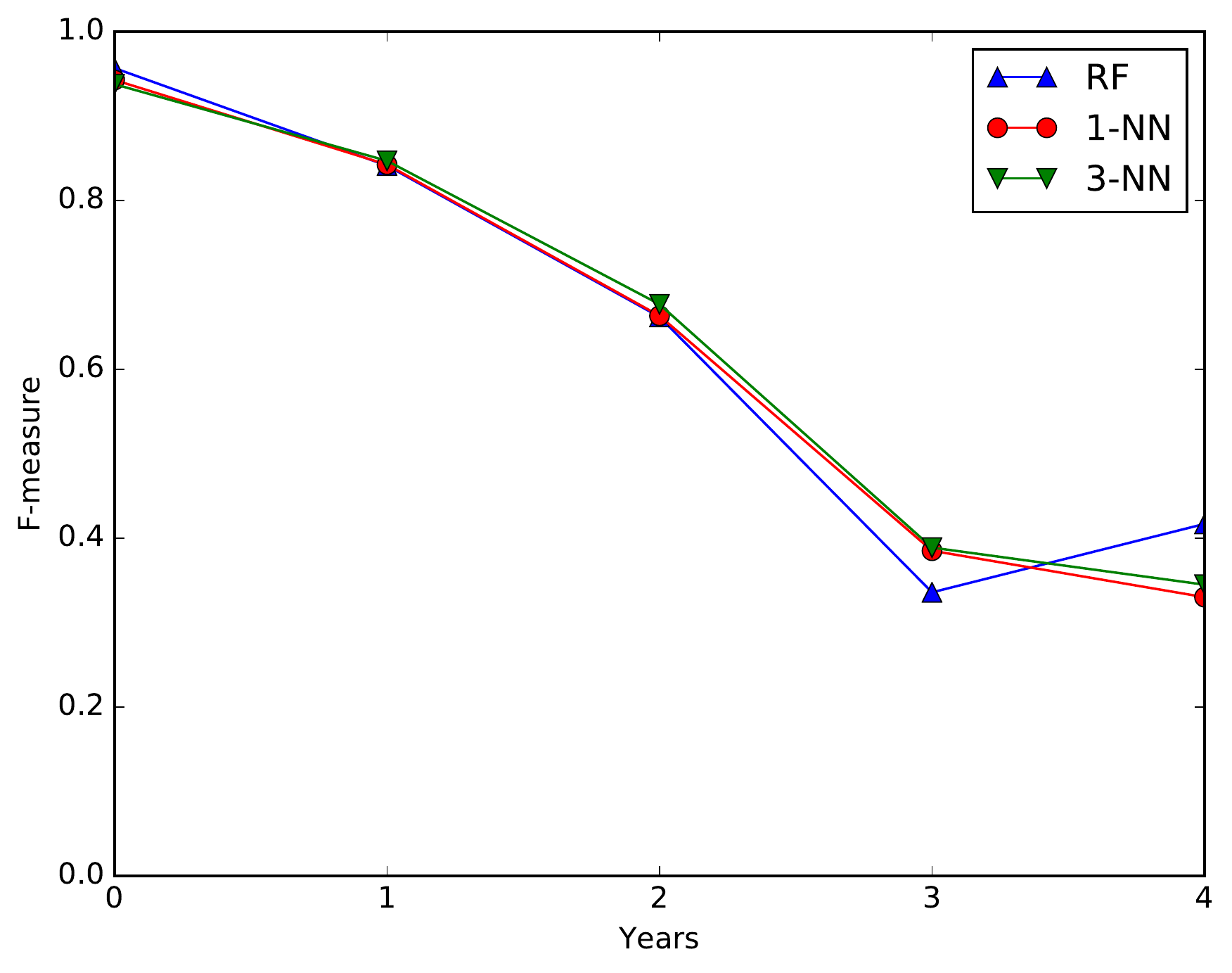} %
        \vspace{-0.2cm}
	\caption{F-measure of \approach classification using older samples for training and newer for
	testing (package mode).}
    \label{fig:FMFutpac}
        \vspace{-0.2cm}
\end{figure}

To verify this hypothesis, we test \approach using older samples as training sets and newer ones as test sets.
\figurename~\ref{fig:FMFutfam} reports the F-measure of
the classification in this setting, with \approach operating in
family mode. The x-axis reports the difference in years between training and test
data. We obtain 86\% F-measure when we classify
apps one year older than the samples on which we train. Classification is still relatively accurate, at 75\%, even after two years.
Then, from
\figurename~\ref{fig:FMFutpac}, we observe that the F-measure does not significantly change when operating in package
mode. Both modes of operations are affected by one particular condition, already discussed in Section
\ref{sec:data}: in our models, benign datasets seem to ``anticipate'' malicious ones by 1--2 years in the way they use certain API calls. 
As a result, we notice a drop in accuracy when classifying future samples and using {\tt drebin} (with samples from 2010 to 2012) or {\tt 2013} as the malicious training set and {\tt oldbenign} (late 2013/early 2014) as the benign training set.
More specifically, we observe that \approach correctly detects benign apps,
while it starts missing true positives and increasing false negatives --- i.e., achieving lower recall.

\begin{figure}[t]
\centering
        \includegraphics[width=0.405\textwidth]{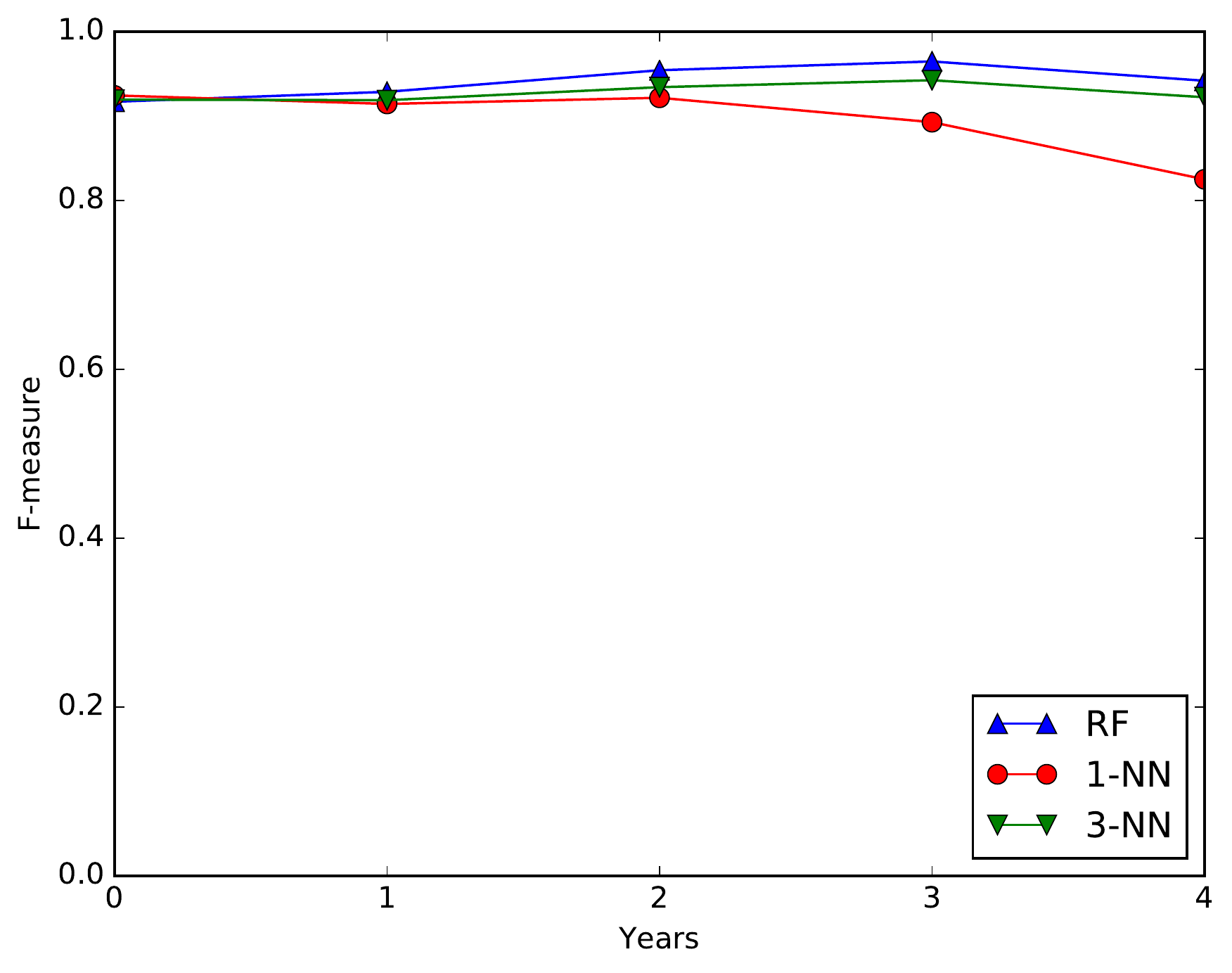} %
        \vspace{-0.2cm}
	\caption{F-measure of \approach classification using newer samples for training and older for testing (family mode).}
    \label{fig:FMPastfam}
        \vspace{-0.2cm}
\end{figure}

\begin{figure}[t]
\centering
        \includegraphics[width=0.405\textwidth]{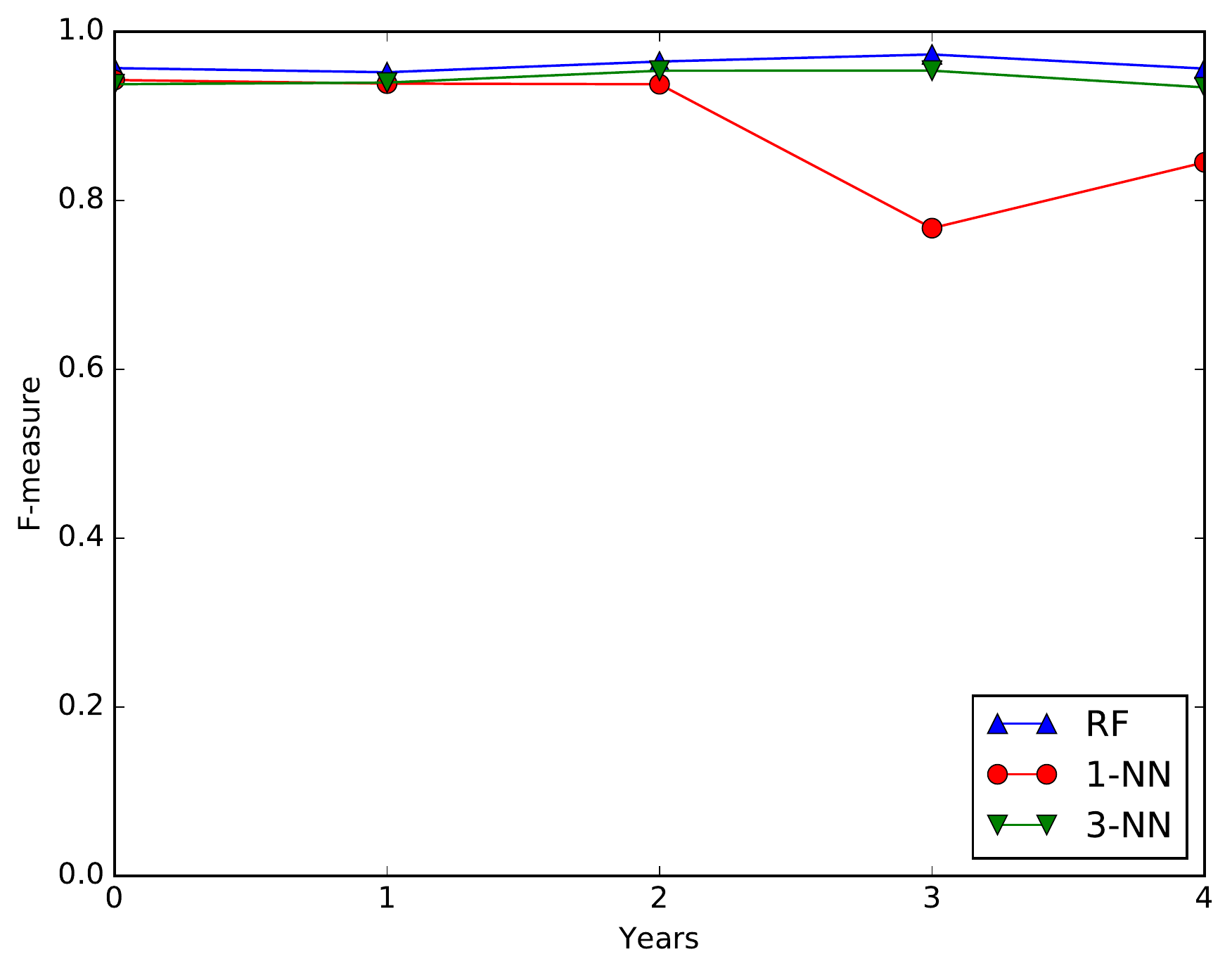} %
        \vspace{-0.2cm}
	\caption{F-measure of \approach classification using newer samples for training and older for testing (package mode).}
    \label{fig:FMPastpac}
        \vspace{-0.2cm}
\end{figure}

We also set to verify whether older malware samples can still be detected
by the system---if not, this would obviously become vulnerable to older (and possibly popular) attacks. 
Therefore, we also perform the ``opposite'' experiment, i.e., training \approach
with newer datasets, and checking whether it is able to detect malware developed years
before. Specifically, \figurename~\ref{fig:FMPastfam} and~\ref{fig:FMPastpac} report results when training
\approach with samples from a given year, and testing it with others that are up to 4 years older:  
\approach retains similar F-measure scores over the years. Specifically, in family mode, 
it varies from 93\% to 96\%, whereas, in package mode, from 95\% to 97\% with the oldest samples. 

\subsection{Case Studies of False Positives and Negatives}
The experiment analysis presented above show that \approach detects Android
malware with high accuracy. As in any detection system, however, the system makes a small number of incorrect classifications,
incurring some false positives and false negatives.
Next, we discuss a few case studies aiming to better understand these misclassifications.
We focus on the experiments with newer datasets, i.e., \texttt{2016} and \texttt{newbenign}.

\descr{False Positives.} We analyze the manifest of the 164 apps
mistakenly detected as malware by \approach, finding that most of them use ``dangerous'' 
permissions~\cite{andriotis2016permissions}. In particular, 67\% of the apps write to external storage, 32\% read the phone state, and 21\% access the device's fine location. We further analyzed %
apps (5\%) that use the READ\_SMS and SEND\_SMS permissions, i.e., even though they
are not SMS-related apps, they can read and send SMSs
as part of the services they provide to users. In particular, a {\em ``in case of emergency''} app is
able to send messages to several contacts from its database (possibly added by the user), which
is a typical behavior of Android malware in our dataset, ultimately leading \approach
to flag it as malicious.

\descr{False Negatives.} We also check the 114 malware samples missed
  by \approach when operating in family mode, using VirusTotal.\footnote{\url{https://www.virustotal.com}} We find that 18\%
  of the false negatives are actually not classified as malware by any of the antivirus engines used
  by VirusTotal, suggesting that these are actually legitimate apps  mistakenly included in
  the VirusShare dataset. 45\% of \approach's false negatives are {\em adware}, typically,
  repackaged apps in which the advertisement library has been substituted with a third-party one,
  which creates a monetary profit for the developers. Since they are not performing any clearly malicious activity,
  \approach is unable to identify them as malware. 
  Finally, we find that 16\% of the false negatives reported by \approach are samples sending text messages or
starting calls to premium services. We also do a similar analysis of false negatives
when abstracting to packages (74 samples), with similar results: there a few more adware samples (53\%), but similar
percentages for potentially benign apps (15\%) and samples sending SMSs or placing calls (11\%).

In conclusion, we find that \approach's sporadic misclassifications are
typically due to benign apps behaving similarly to malware,
malware that do not perform clearly-malicious activities, or 
mistakes in the ground truth labeling.

\begin{table*}[t]
\centering
\small
\setlength{\tabcolsep}{3pt}
\resizebox{0.999\linewidth}{!}{
\begin{tabular}{|c|cc|cc|cc|cc|cc|}
\cline{2-11}
\multicolumn{1}{c|}{} & \multicolumn{10}{c|}{\bf Testing Sets}   \\
\cline{2-11}
\multicolumn{1}{c|}{} & \multicolumn{2}{c|}{\texttt{drebin} \& {\tt oldbenign}} & \multicolumn{2}{c|}{\texttt{2013} \& {\tt oldbenign}} & \multicolumn{2}{c|}{\texttt{2014} \& {\tt oldbenign}} & \multicolumn{2}{c|}{\texttt{2015} \& {\tt oldbenign}} & \multicolumn{2}{c|}{\texttt{2016} \& {\tt oldbenign}}\\
\hline %
{\bf Training Sets} & \hspace*{0.4cm} \cite{Aafer2013DroidAPIMiner} \hspace*{0.4cm} & Our Work & \hspace*{0.6cm} \cite{Aafer2013DroidAPIMiner} \hspace*{0.6cm}  & Our Work & \hspace*{0.6cm}  \cite{Aafer2013DroidAPIMiner} \hspace*{0.6cm} & Our Work & \hspace*{0.6cm}  \cite{Aafer2013DroidAPIMiner} \hspace*{0.6cm} & Our Work &  \hspace*{0.6cm} \cite{Aafer2013DroidAPIMiner} \hspace*{0.6cm} & Our Work\\
\hline
\texttt{drebin \& oldbenign} & 0.32 & {\bf 0.96} & 0.35 & {\bf 0.95} & 0.34 & {\bf 0.72} & 0.30 & {\bf 0.39} & 0.33 & {\bf 0.42} \\
\hline
\texttt{2013 \& oldbenign} & 0.33 & {\bf 0.94} &  0.36  & {\bf 0.97} & 0.35 & {\bf 0.73} & 0.31 & {\bf 0.37} & {\bf 0.33} & 0.28 \\
\hline
\texttt{2014 \& oldbenign} & 0.36 & {\bf 0.92} & 0.39 & {\bf 0.93} & 0.62 & {\bf 0.95} & 0.33 & {\bf 0.78} & 0.37 & {\bf 0.75} \\
\hline
 & \multicolumn{2}{c|}{\texttt{drebin} \& {\tt newbenign}} & \multicolumn{2}{c|}{\texttt{2013} \& {\tt newbenign}} & \multicolumn{2}{c|}{\texttt{2014} \& {\tt newbenign}} & \multicolumn{2}{c|}{\texttt{2015} \& {\tt newbenign}} & \multicolumn{2}{c|}{\texttt{2016} \& {\tt newbenign}}\\
\hline 
{\bf Training Sets} & \hspace*{0.4cm} \cite{Aafer2013DroidAPIMiner} \hspace*{0.4cm} & Our Work & \hspace*{0.6cm} \cite{Aafer2013DroidAPIMiner} \hspace*{0.6cm}  & Our Work & \hspace*{0.6cm}  \cite{Aafer2013DroidAPIMiner} \hspace*{0.6cm} & Our Work & \hspace*{0.6cm}  \cite{Aafer2013DroidAPIMiner} \hspace*{0.6cm} & Our Work &  \hspace*{0.6cm} \cite{Aafer2013DroidAPIMiner} \hspace*{0.6cm} & Our Work\\
\hline
\texttt{2014 \& newbenign} & 0.76 & {\bf 0.98} & 0.75 & {\bf 0.98} & 0.92 & {\bf 0.99} & 0.67 & {\bf 0.85} & 0.65 & {\bf 0.81} \\
\hline
\texttt{2015 \& newbenign} & 0.68 & {\bf 0.97} & 0.68 & {\bf 0.97} & 0.69 & {\bf 0.99} & 0.77 & {\bf 0.95} & 0.65 & {\bf 0.88} \\
\hline
\texttt{2016 \& newbenign} & 0.33 & {\bf 0.96} & 0.35 & {\bf 0.98} & 0.36 & {\bf 0.98} & 0.34 & {\bf 0.92} & 0.36 & {\bf 0.92} \\
\hline
\end{tabular}
}
\caption{Classification performance of \droid~\cite{Aafer2013DroidAPIMiner} vs \approach (our work).}
\vspace{-0.35cm}
\label{table:droidapiresults}
\end{table*}

\subsection{\approach vs \droid}\label{sec:compare}
We also compare the performance of \approach 
to previous work using API features for Android malware classification.
Specifically, we compare to \droid~\cite{Aafer2013DroidAPIMiner},
because: (i) it uses API calls and its parameters to perform classification;
(ii) it reports high true positive rate (up to 97.8\%) on almost 4K malware samples obtained from McAfee and {\sc Genome}~\cite{Zhou2012dissecting}, and
16K benign samples; and (iii) its source code has been made available to us by the authors. 

In \droid, permissions that are requested more frequently
by malware samples than by benign apps are used to perform a baseline classification. 
Since there are legitimate situations where a non-malicious app needs permissions tagged as dangerous,
\droid also applies frequency analysis on the list of
API calls, specifically, using the 169 most frequent API calls in the malware samples (occurring at least 6\% more in malware than benign samples) ---leading to a reported 83\% precision. 
Finally, data flow analysis is applied on the API calls that are frequent in both
benign and malicious samples, but do not occur by at least, 6\% more in the malware set. Using the top 60 parameters, 
the 169 most frequent calls change, and authors report a precision of 97.8\%. 

After obtaining \droid's source code, as well as a list of packages used for feature refinement,
we re-implement the system by modifying the code in order to reflect recent changes in Androguard (used by \droid for API call extraction),
extract the API calls for all apps in the datasets listed in Table~\ref{table:dataset}, and perform a frequency
analysis on the calls.  Androguard fails to extract calls for about 2\% (1,017)
of apps in our datasets as a result of bad CRC-32 redundancy checks and error in
unpacking, thus \droid is evaluated over the samples in the second-to-last column of Table~\ref{table:dataset}. 
We also implement classification, which is missing from the code
provided by the authors, using k-NN (with k$=$3) since it achieves the best results according to the paper. We use 2/3 of the dataset for training and 1/3 for testing as implemented by the authors~\cite{Aafer2013DroidAPIMiner}.
A summary of the resulting F-measures %
obtained using different training and test sets is presented in Table~\ref{table:droidapiresults}. 

We set up a number of experiments to thoroughly compare \droid to \approach.
First, we set up three experiments in which we train \droid using a dataset composed of {\tt
oldbenign} combined with one of the three
oldest malware datasets each ({\tt drebin}, {\tt 2013}, and {\tt 2014}), and testing on all malware datasets. 
With this configuration, the best result (with
{\tt 2014} and {\tt oldbenign} as training sets) amounts to 62\% F-measure when tested on
the same dataset. The F-measure drops to 33\% and 39\%, respectively, when tested on samples one year into the
future and past. If we use the same configurations in \approach, 
in package mode, we obtain up to 97\% F-measure (using {\tt 2013} and
{\tt oldbenign} as training sets), dropping to 73\% and 94\%, respectively, one year into the future and into the past.
For the datasets where \droid achieves its best result (i.e., {\tt 2014} and {\tt
oldbenign}), \approach achieves an F-measure of 95\%, which drops to respectively, 78\% and 93\%
one year into the future and the past. The F-measure is stable even two years into the future and the past at 75\% and 92\%, respectively. 

As a second set of experiments, we train \droid using a dataset composed of {\tt newbenign} combined
with one of the three most recent malware datasets each ({\tt 2014}, {\tt 2015}, and {\tt 2016}).
Again, we test \droid on all malware datasets. The best result is obtained with the dataset ({\tt
2014} and {\tt newbenign}) used for both testing and training, yielding a
F-measure of 92\%, which drops to 67\% and 75\% one year into the future and past respectively.
Likewise, we use the same datasets for \approach, with the best
results achieved on the same dataset as \droid. In package mode, \approach achieves an F-measure 
of 99\%, which is maintained more than two years into the past, but drops to respectively, 85\% and 81\% one and two years into the future. 

As summarized in Table~\ref{table:droidapiresults}, \approach achieves significantly higher performance than \droid in all but one experiment, %
with the F-measure being at least 75\% even after two years into the future or the past when datasets from 2014 or later are used for training. 
Note that there is only one setting in which \droid performs slightly better than
\approach: this occurs when the malicious training set is much older than the malicious test set.
Specifically, \approach presents low recall in this case: as discussed, \approach's classification performs much better when the training set is not more than two years older than the test set.

\subsection{Runtime Performance}
\label{sub:timeeval}
We envision \approach to be integrated in offline detection systems, e.g., run by Google Play. 
Recall that \approach consists of different phases, so in the following, we review the computational overhead incurred by each of them, aiming to assess the feasibility of real-world deployment. We run our experiments on a desktop  
equipped with an 40-core 2.30GHz CPU and 128GB of RAM, but only use one core and allocate 16GB of RAM for evaluation. %

\approach's first step involves extracting the call graph from an apk and the complexity of this task varies significantly across apps. On average, it takes 9.2s$\pm$14 (min 0.02s, max 13m)
to complete for samples in our malware sets. %
Benign apps usually yield larger call graphs, and the average time to extract them is 25.4s$\pm$63 (min 0.06s, max 18m) 
per app. %
Note that we do not include in our evaluation apps for which we could not successfully extract the call graph. 

Next, we measure the time needed to extract call sequences while abstracting to
families or packages, depending on \approach's mode of operation. In family
mode, this phase completes in about 1.3s on average (and at most 11.0s) with both benign and malicious samples. 
Abstracting to packages takes slightly longer, due to the use of 341 packages in
\approach. On average, this extraction takes 1.67s$\pm$3.1 for malicious apps and
1.73s$\pm$3.2 for benign samples. As it can be seen, the call sequence extraction in package
mode does not take significantly more than in family mode.

\approach's third step includes Markov chain modeling and feature vector
extraction. This phase is fast regardless of the mode of operation and datasets
used. Specifically, with malicious samples, it takes on average 0.2s$\pm$0.3 and 2.5s$\pm$3.2 (and at most 2.4s and 22.1s), respectively, with families and packages, whereas, with benign samples, averages rise to 0.6s$\pm$0.3 and 6.7s$\pm$3.8 (at most 1.7s and 18.4s).

Finally, the last step involves classification, and performance depends on both the machine learning algorithm employed
and the mode of operation. More specifically, running times are affected by the number of features for the app to be classified, 
and not by the initial dimension of the call graph, or by whether the app is benign or malicious. 
Regardless, in family mode, Random Forests, 1-NN, and 3-NN all take less than 0.01s. With packages, it takes, respectively, 0.65s, 1.05s, and 0.007s per app with 1-NN, 3-NN, Random Forests. 

Overall, when operating in family mode, malware and benign samples take on average, 10.7s and 27.3s respectively to complete the entire process, from call graph extraction to classification.  %
Whereas, in package mode, the average completion times for malware and benign samples are 13.37s and 33.83s respectively. In both modes of operation, time is mostly (> 80\%) spent on call graph extraction. %

We also evaluate the runtime performance of \droid~\cite{Aafer2013DroidAPIMiner}. Its first step, i.e., extracting API calls, takes 0.7s$\pm$1.5 (min 0.01s, max 28.4s) per app in our malware datasets. %
Whereas, it takes on average 13.2s$\pm$22.2 (min 0.01s, max 222s) per benign app. %
In the second phase, i.e., frequency and data flow analysis, it takes, on average, 4.2s per app. Finally, classification using 3-NN is very fast: 0.002s on average. Therefore, in total, \droid takes respectively, 17.4s and 4.9s for a complete execution on one app from our benign and malware datasets, %
which while faster than \approach, %
achieves significantly lower accuracy.

In conclusion, our experiments show that our prototype implementation of \approach is scalable enough to be deployed. Assuming that, everyday, a number of apps in the order of 10,000 are submitted to Google Play, and using the average execution time of benign samples in family (27.3s) and package (33.83s) modes, we estimate that it would take less than an hour and a half to complete execution of all apps submitted daily in both modes, with just 64 cores. Note that we could not find accurate statistics reporting the number of apps submitted everyday, but only the total number of apps on Google Play.\footnote{\url{http://www.appbrain.com/stats/number-of-android-apps}} On average, this number increases of a couple of thousands per day, and although we do not know how many apps are removed, we believe 10,000 apps submitted every day is likely an upper bound. %

\section{Discussion}
\label{sec:discussion}

We now %
discuss the implications of our results with
respect to the feasibility of modeling app behavior using static analysis and Markov chains, 
discuss possible evasion techniques, 
and highlight some limitations of our approach.

\subsection{Lessons Learned}
Our work yields important insights around the use of API calls in malicious apps, showing that, by modeling the sequence of API calls made by an app as a Markov chain, we can successfully capture the behavioral model of that app. 
This allows \approach to obtain high accuracy overall, as well as to retain it over the years,
which is crucial due to the continuous evolution of the Android ecosystem.

As discussed in Section~\ref{sec:data}, the use of API calls changes over time, and in different ways across malicious and benign samples. 
From our newer datasets, which include samples up to Spring 2016 (API level 23), we observe that
newer APIs introduce more packages, classes, and methods, while also deprecating some. %
\figurename~\ref{fig:NumCalls},~\ref{fig:FamAndro}, and~\ref{fig:FamGoogle} show that benign apps are using more
calls than malicious ones developed around the same time. We also notice an interesting trend in
the use of Android and Google APIs: malicious apps follow the same trend as
benign apps in the way they adopt certain APIs, but with a delay of some years. %
This might be a side effect of Android malware authors' tendency
to repackage benign apps, adding their malicious functionalities onto them.

Given the frequent changes in the Android framework and the continuous evolution of malware, systems like \droid~\cite{Aafer2013DroidAPIMiner} -- being dependent on the presence or the use
of certain API calls -- become increasingly less effective with time. %
As shown in Table~\ref{table:droidapiresults}, %
malware that uses API calls released after those used by samples in the training set cannot be identified by these systems.
On the contrary, as shown in \figurename~\ref{fig:FMFutfam} and~\ref{fig:FMFutpac}, \approach detects malware samples that are {\em 1 year} newer than the training set obtaining an 86\% F-measure (as opposed to 46\% with \droid). After 2 years, the value is still at 75\% (42\% with \droid), dropping to 51\% after 4 years.

We argue that the effectiveness of \approach's classification remains relatively high ``over the years'' owing to
Markov models capturing app behavior. These models tend to be more robust to malware evolution
because abstracting to families or packages makes the system less susceptible to the introduction of new API calls.
Abstraction allows \approach to capture newer
classes/methods added to the API, since these are abstracted to already-known families or packages.
In case newer packages are added to the API, and these packages start being used by malware,
\approach only requires adding a new state to the Markov chains, and probabilities of a transition from a state to this new state in old apps would be 0. Adding only a few nodes does not likely alter the probabilities of the other 341 nodes, thus, two apps created with the same purpose will not strongly differ in API calls usage if they are developed using almost consecutive API levels.

We also observe that abstracting to packages provides a slightly better tradeoff than
families. In family mode, the system is lighter and faster, and actually performs better 
when there are more than two years between training and test set samples
However, even though both modes of operation effectively detect malware, abstracting to packages yields better results overall.
Nonetheless, this does not imply that less
abstraction is always better: in fact, a system that is too granular, besides incurring untenable
complexity, would likely create Markov models with low-probability transitions, ultimately resulting
in less accurate classification.
We also highlight that applying PCA is a good strategy to preserve high 
accuracy and at the same time reducing complexity.
\subsection{Evasion}
Next, we discuss possible evasion techniques and how they can be addressed.
One straightforward evasion approach could be to repackage a benign app with small snippets of malicious code added to a few classes. However, it is difficult to embed malicious code in such a way that, at the same time, the resulting Markov
chain looks similar to a benign one. For instance, our running example from Section~\ref{sec:method} 
(malware posing as a memory booster app and executing unwanted commands as root) 
is correctly classified by \approach; although most functionalities in
this malware are the same as the original app, injected API calls generate some transitions in the Markov chain that are not typical of benign samples. %

The opposite procedure -- i.e., embedding portions of benign code into a malicious app -- is also likely ineffective against \approach, since, for each app, we derive the feature vector from the transition probability between calls over the entire app. In other words, a malware developer would have to embed benign code inside the malware in such a way that the overall sequence of calls yields similar transition probabilities as those in a benign app, but this is difficult to achieve because if the sequences of calls have to be different (otherwise there would be no attack), then the models will also be different. 

An attacker could also try to create an app from scratch with a similar Markov
chain to that of a benign app. Because this is derived from the sequence of
abstracted API calls in the app, it is actually very difficult to create sequences resulting in Markov chains similar to benign apps while, at the same time, actually engaging in malicious behavior. Nonetheless, in future work, we plan to systematically analyze the feasibility of this strategy.

Moreover, attackers could try using reflection, dynamic code loading, or native code~\cite{poeplau2014execute}. Because \approach uses static analysis, 
it fails to detect malicious code when it is loaded or determined at runtime.
However, \approach can detect reflection when a method from the reflection
package ({\tt java.lang.reflect}) is executed. Therefore, we obtain the correct
sequence of calls up to the invocation of the reflection call, which may be
sufficient to distinguish between malware and benign apps. Similarly, \approach
can detect the usage of class loaders and package contexts that can be used to
load arbitrary code, but it is not able to model the code loaded; likewise, native code that is part of the app cannot be modeled, as it is not Java and is not processed by Soot. These limitations are not specific of \approach, but are a problem of static analysis in general, which can be mitigated by using \approach alongside dynamic analysis techniques.

Malware developers might also attempt to evade \approach by naming their self-defined packages in such a way that they look similar to that of the {\tt android}, {\tt java}, or {\tt google} APIs, e.g., creating packages like java.lang.reflect.{\em malware} and java.lang.{\em malware}, aiming to confuse \approach into abstracting them to respectively, {\tt java.lang.reflect} and {\tt java.lang}. However, this is easily prevented by whitelisting the list of packages from {\tt android}, {\tt java}, or {\tt google} APIs. %

Another approach could be using dynamic dispatch so that a class X in package A is created to extend class Y in package B with static analysis reporting a call to root() defined in Y as X.root(), whereas, at runtime Y.root() is executed. 
This can be addressed, however, with a small increase in \approach's computational cost, by keeping track of self-defined classes that extend or implement classes in the recognized APIs, and abstract polymorphic functions of this self-defined class to the corresponding recognized package, while, at the same time, abstracting as self-defined overridden functions in the class. %

Finally, identifier mangling and other forms of obfuscation could be used aiming to obfuscate code and hide malicious actions. However, since classes in the Android framework cannot be obfuscated by obfuscation tools, malware developers can only do so for self-defined classes. \approach labels obfuscated calls as {\tt obfuscated} so, ultimately, these would be captured in the behavioral model (and the Markov chain) for the app. %
In our sample, we observe that benign apps use significantly less obfuscation than malicious apps, indicating that obfuscating a significant number of classes is not a good evasion strategy since this would likely make the sample more easily identifiable as malicious.

\subsection{Limitations}\label{sec:limits}
\approach requires a sizable amount of memory in order to perform classification, when operating in package mode, working on more than 100,000 features per sample. 
The quantity of features, however, can be further reduced using feature selection algorithms such as
PCA. As explained in Section \ref{sec:evaluation} when we use 10 components from the PCA the system performs almost as well as the one using all the features; however, using PCA comes with a
much lower memory complexity in order to run the machine learning algorithms, because the number of dimensions of the features space where the classifier operates is remarkably reduced.

Soot~\cite{ValleeRai1999Soot}, which we use to extract call graphs, fails to analyze some apks. In fact, %
we were not able to extract call graphs for a fraction (4.6\%) %
of the apps in the original datasets due to scripts either %
failing to apply the {\tt jb} phase, which is used to transform Java bytecode to the primary intermediate representation (i.e., jimple) of Soot or not able to open the apk. %
Even though this does not really affect the results of our evaluation, one could avoid it
by using %
a different/custom intermediate representation for the analysis or use different tools to extract the call graphs. 

In general, static analysis methodologies for malware detection on Android could fail to capture the runtime
environment context, code that is executed more frequently, or other effects stemming from user input~\cite{arp2014drebin}.
These limitations can be addressed using dynamic analysis, or by recording function calls
on a device. %
Dynamic analysis observes the live performance of the samples, recording what activity is actually
performed at runtime. Through dynamic analysis, it is also possible to provide inputs to the app and
then analyze the reaction of the app to these inputs, going beyond static analysis limits. To this end, we plan
to integrate dynamic analysis to build the models used by \approach as part of future work.

\section{Related Work}
\label{sec:related}

Over the past few years, Android security has attracted a wealth of work by
the research community. In this section, we review (i) program analysis
techniques focusing on general security properties of Android apps, and then
(ii) systems that specifically target malware on Android.\vspace{-0.15cm}

\subsection{Program Analysis} 
Previous work on program analysis applied to Android security has used both
static and dynamic analysis. With the former, the program's code is decompiled in order to extract features without actually running the program, usually employing tools such as Dare~\cite{Octeau2012dare} to obtain Java bytecode. The latter involves real-time execution of the program, typically in an emulated or protected environment.

Static analysis techniques include work by Felt et al.~\cite{Felt2011}, who analyze API calls to
identify over-privileged apps, while Kirin~\cite{Enck2009} is a system that examines permissions requested by apps to perform a lightweight certification, using a set of security rules that indicate whether or not the security configuration bundled with the app is safe.
RiskRanker~\cite{Grace2012riskranker} aims to identify zero-day Android malware
by assessing potential security risks caused by untrusted apps. It sifts through a large number of apps from Android markets and examines them to detect certain behaviors, such as encryption and dynamic code loading, which form malicious patterns and can be used to detect stealthy malware. Other methods, such as CHEX~\cite{Lu2012chex}, use
data flow analysis to automatically vet Android apps for
vulnerabilities. %
Static analysis has also been applied to the detection of data leaks and malicious data flows from Android apps~\cite{Arzt2014flowdroid, Klieber2014, Yang2013app, kim2012scandal}. 

DroidScope~\cite{Yan2012droidscope} and TaintDroid~\cite{Enck2014taintdroid}
monitor run-time app behavior in a protected environment %
to perform dynamic taint analysis. DroidScope performs dynamic taint analysis at the machine code level, while TaintDroid  monitors how third-party apps access or manipulate users' personal data, aiming to detect sensitive data leaving the system. However, as it is unrealistic to deploy dynamic analysis techniques directly on users' devices, due to the overhead they introduce, these are typically used offline~\cite{Rastogi2013, Zhou2012hey, tam2015copperdroid}. ParanoidAndroid~\cite{Portokalidis2010} employs a virtual clone of the smartphone, %
running in parallel in the cloud and replaying activities of the device -- however, even if minimal execution traces are actually sent to the cloud, this still takes a non-negligible toll on battery life.

Recently, hybrid systems like IntelliDroid~\cite{wong2016intellidroid} have also been proposed that use input generators, producing inputs specific to dynamic analysis tools. Other work combining static and dynamic analysis include~\cite{Ge2011, jiang2013detecting, Xia2015, Bhoraskar2014}.

\subsection{Android Malware Detection} 
A number of techniques have used {\em signatures} for Android malware detection. 
NetworkProfiler~\cite{Dai2013} generates network profiles for Android apps and extracts fingerprints based on such traces, 
while work in~\cite{Canfora2016} obtains resource-based metrics (CPU, memory, storage, network) to distinguish malware activity from benign one.
Chen et al.~\cite{Chen2016storm} extract statistical features, such as
permissions and API calls, and extend their vectors to add dynamic
behavior-based features. While their experiments show that their %
solution outperforms, in terms of accuracy, other antivirus systems, Chen et al.~\cite{Chen2016storm} indicate that the quality of their detection model critically depends on the availability of representative benign and malicious apps for training.
Similarly, ScanMe Mobile~\cite{Zhang2016} uses the Google Cloud Messaging Service (GCM) 
to perform static and dynamic analysis on apks found on the device's SD card. %

The sequences of system calls have also been used to detect malware in both desktop and Android environments.
Hofmeyr et al.~\cite{hofmeyr1998intrusion} demonstrate that
short sequences of system calls can be used as a signature to discriminate
between normal and abnormal behavior of common UNIX programs. Signature-based
methods, however, can be evaded using polymorphism and obfuscation, as well as
by call re-ordering attacks~\cite{kolbitsch2009effective}, even though
quantitative measures, such as similarity analysis, can be used to address some
of these attacks~\cite{Shankarapani2011}. \approach inherits the spirit of these
approaches, proposing a statistical method to model app behavior that is
more robust against evasion attempts.

In the Android context, Canfora et al.~\cite{Canfora2015} %
use the sequences of three system calls (extracted from the execution traces of
apps under analysis) to detect malware. %
This approach models specific malware families, aiming to identify additional
samples belonging to such families. %
In contrast, \approach's goal is to detect previously-unseen malware, and we
also show that our system can detect new malware families that even appear years after the
system has been trained. %
In addition, using strict sequences of system or API calls can be easily evaded
by malware authors who could add unnecessary calls %
to effectively evade detection. Conversely, \approach builds a behavioral model of an Android app, 
which makes it robust to this type of evasion. %

Dynamic analysis has also been applied to detect Android malware by using predefined scripts of common inputs that will be performed when the device is running. However, this might be inadequate due to the low probability of triggering malicious behavior, and can be side-stepped by knowledgeable adversaries, as suggested by Wong and Lie~\cite{wong2016intellidroid}. Other approaches include random fuzzing~\cite{Machiry2013, Ye2013} and concolic testing~\cite{Anand2012, Godefroid2005}. Dynamic analysis can only detect malicious activities if the code exhibiting malicious behavior is actually running during the analysis. Moreover, according to~\cite{Vidas2014}, mobile malware authors often employ emulation or virtualization detection strategies to change malware behavior and eventually evade detection.

Aiming to complement static and dynamic analysis tools, machine learning techniques have also been applied to assist Android malware detection. %
Droidmat~\cite{Wu2012Droidmat} uses API call tracing and manifest files to learn features for malware detection, while Gascon et al.~\cite{Gascon2013} rely on embedded call graphs. DroidMiner~\cite{yang2014droidminer} studies the program logic of sensitive Android/Java framework API functions and resources, and detects malicious behavior patterns. MAST~\cite{Chakradeo2013} statically analyzes apps using features such as permissions, presence of native code, and intent filters and measures the correlation between multiple qualitative data. %
Crowdroid~\cite{Burguera2011Crowdroid} relies on crowdsourcing to distinguish between malicious and benign apps by monitoring system calls. 
AppContext~\cite{Yang2015appcontext} models security-sensitive behavior, such as activation events or environmental attributes, and uses SVM  to classify these behaviors, while RevealDroid~\cite{garcia2015obfuscation} employs supervised learning and obfuscation-resilient methods targeting API usage and intent actions to identify their families. 

{\sc Drebin}~\cite{arp2014drebin} automatically deduces detection patterns and
identifies malicious software directly on the device, performing a broad static
analysis. This is achieved by gathering numerous features from the manifest file
as well as the app's source code (API calls, network addresses, permissions). 
Malevolent behavior is reflected in
patterns and combinations of extracted features from the static analysis: for
instance, the existence of both SEND\_SMS %
permission and the android.hardware.telephony component in an app might indicate an attempt 
to send premium SMS messages, and this combination can eventually
constitute a detection pattern. %

In Section~\ref{sec:evaluation}, we have already introduced, and compared against,
\droid~\cite{Aafer2013DroidAPIMiner}. This system relies on the top-169 API calls that are used more frequently in the malware than in the benign set, along with data flow analysis
on calls that are frequent in both benign and malicious apps, but occur up to
6\% more in the latter. As shown in our evaluation, using the most common calls observed
during training requires constant retraining, due to the evolution of both
malware and the Android API. On the contrary, \approach can effectively model
both benign and malicious Android apps, and perform an efficient classification
on them. Compared to \droid, our approach is more resilient to changes in the
Android framework than \droid, resulting in a less frequent need to re-train the classifier. %

Overall, compared to state-of-the-art systems like {\sc Drebin}~\cite{arp2014drebin} and \droid~\cite{Aafer2013DroidAPIMiner}, \approach is more generic and robust as its statistical
modeling does not depend on specific app characteristics, but can actually be run on any app created for any Android API level.

Finally, also related to \approach are Markov-chain based models for Android
malware detection. Chen et al.~\cite{Chen2014Hmm} dynamically analyze 
system- and developer-defined actions from intent messages (used by app components to communicate with each other at runtime),
and probabilistically estimate whether an app is performing benign or malicious actions at run time, but 
obtain low accuracy overall. Canfora et al.~\cite{Canfora2016HMM} use a Hidden Markov model (HMM) to
identify malware samples belonging to previously observed malware families, whereas, 
\approach can detect previously unseen malware, not relying on specific malware families. 

\section{Conclusion}
\label{sec:conclusion}

This paper presented \approach, an Android malware detection system based on modeling the sequences
of API calls as Markov chains.
Our system is designed to operate in one of two modes, with different granularities,
by abstracting API calls to either families or packages. 
We ran an extensive experimental evaluation using, to the best of our knowledge, the largest malware dataset ever
analyzed in an Android malware detection research paper, and aiming at assessing both the accuracy of the classification 
(using F-measure, precision, and recall) and runtime performances.
We showed that \approach effectively detects unknown malware samples developed earlier or around the same time
as the samples on which it is trained (F-measure up to 99\%).
It also maintains good detection performance: one year after the model has been trained the F-measure value is 87\%, and after two years it is 73\%.

We compared \approach to \droid~\cite{Aafer2013DroidAPIMiner}, a
state-of-the-art system based on API calls frequently used by malware, showing that, not only does \approach outperforms \droid when trained and tested on the same
datasets, but that it is also much more resilient over the years to changes in the Android API. %
Overall, our results demonstrate that %
the type of statistical behavioral models introduced by \approach are more robust than
traditional techniques, highlighting how our work can form the basis of more advanced detection systems in the future.
As part of future work, we plan to further investigate the resilience to possible evasion techniques, focusing on repackaged malicious apps
as well as injection of API calls to maliciously alter Markov models. We also plan to explore the use of finer-grained abstractions as well as
the possibility to seed the behavioral modeling performed by \approach with dynamic instead of static analysis. Due to the large size of the data, we have not made them readily available online but both the datasets and the feature vectors can be obtained upon request.

\descr{Acknowledgments.} We wish to thank the anonymous reviewers for their
feedback, our shepherd Amir Houmansadr for his
help in improving our paper, and Yousra Aafer for kindly sharing the \droid
source code with us. We also wish to thank Yanick Fratantonio for his
comments on an early draft of the paper. This research was supported by the
EPSRC under grant EP/N008448/1, by an EPSRC-funded ``Future Leaders in
Engineering and Physical Sciences''
award, a Xerox University Affairs Committee grant, and by a small grant from GCHQ. Enrico Mariconti was supported by the
EPSRC under grant 1490017, while Lucky Onwuzurike was funded by the Petroleum
Technology Development Fund (PTDF).

\bibliographystyle{abbrv}

\end{document}